\documentclass[preprint,review,12pt]{elsarticle}

\usepackage{amsmath,amssymb,amsthm}

\usepackage{caption}
\usepackage{multirow}

\usepackage{algorithm} 
\usepackage{algcompatible}
\usepackage{xcolor}
\usepackage{multicol}
\usepackage{tabularx,calc}

\usepackage{bbm}
\newcommand{\peq}{{p_{eq}}}

\newcommand{\mathi}{\textrm{i}}

\newcommand{\EEA}{\end{eqnarray}}
\newcommand{\BEA}{\begin{eqnarray}}
\newcommand{\comment}[1]{}

\usepackage[bookmarks,colorlinks,breaklinks]{hyperref}
\hypersetup{
    colorlinks,%
    citecolor=black,%
    filecolor=black,%
    linkcolor=black,%
    urlcolor=black
}

\journal{Physica D}

\begin{document}

\begin{frontmatter}

\title{Forecasting Turbulent Modes with Nonparametric Diffusion Models: Learning from Noisy Data}

\author{Tyrus Berry}
\ead{tbh11@psu.edu}
\address{Department of Mathematics, The Pennsylvania State University, University Park, PA 16802}

\author{John Harlim\corref{cor}}
\cortext[cor]{Corresponding author.}
\ead{jharlim@psu.edu}
\address{Department of Mathematics and Department of Meteorology, The Pennsylvania State University, University Park, PA 16802}

\begin{abstract}
In this paper, we apply a recently developed nonparametric modeling approach, the ``diffusion forecast", to predict the time-evolution of Fourier modes of turbulent dynamical systems. While the diffusion forecasting method assumes the availability of a noise-free training data set observing the full state space of the dynamics, in real applications we often have only partial observations which are corrupted by noise. 
To alleviate these practical issues, following the theory of embedology, the diffusion model is built using the delay-embedding coordinates of the data.  We show that this delay embedding biases the geometry of the data in a way which extracts the most stable component of the dynamics and reduces the influence of independent additive observation noise. The resulting diffusion forecast model approximates the semigroup solutions of the generator of the underlying dynamics in the limit of large data and when the observation noise vanishes. As in any standard forecasting problem, the forecasting skill depends crucially on the accuracy of the initial conditions. We introduce a novel Bayesian method for filtering the discrete-time noisy observations which works with the diffusion forecast to determine the forecast initial densities.

Numerically, we compare this nonparametric approach with standard stochastic parametric models on a wide-range of well-studied turbulent modes, including the Lorenz-96 model in weakly chaotic to fully turbulent regimes and the barotropic modes of a quasi-geostrophic model with baroclinic instabilities. We show that when the only available data is the low-dimensional set of noisy modes that are being modeled, the diffusion forecast is indeed competitive to the perfect model.
\end{abstract}


\end{frontmatter}

\section{Introduction}

A long-standing issue in modeling turbulent dynamics is the so-called turbulent closure problem (see e.g.\cite{davidson2004turbulence}) where the goal is to find a set of effective equations to represent low-order statistics of the coarse-grained variables of interest. The main difficulty of this problem is largely due to the infinite dimensionality and nontrivial coupling of the governing equations of the statistics.
In order to predict a few lower-order statistics of some resolved variables, common closure approaches were developed using physical insights to choose a parametric ansatz to represent the feedback from the unresolved scales (see e.g., \cite{kmv:03} for various closure approximations for predicting passive scalar turbulence and \cite{delsole:04,kbm:10,gm:13,gm:14} for various stochastic modeling approaches for predicting geophysical turbulence).

Despite these successes, the parametric modeling approaches have practical issues due to model error when the necessary physical insights are not known. If the parametric model (or ansatz) is not chosen appropriately, one can end up with a model with poor predictive skills (or even with solutions which diverge catastrophically) even when the parameters can be obtained by a standard regression fitting procedure \cite{my:12}. Moreover, even when an appropriate parametric form is chosen, specifying the parameters from noisy observations of the physical variables can be nontrivial since the parameters are typically not directly observed. Indeed, it was shown that an appropriate parameterization scheme is crucial for accurate filtering and equilibrium statistical prediction even when the parametric forms are appropriately chosen \cite{PRSA}. 

Recently, a nonparametric modeling approach, called the \emph{diffusion forecast}, for predicting the evolution of the probability density of low-dimensional dynamical system was introduced in \cite{bgh:15}. The approach of \cite{bgh:15} can be intuitively viewed as extending the standard nonparametric statistical models (such as kernel density estimates) which are used to estimate time-independent densities \cite{semiparamBook}. The key idea behind the diffusion forecast is to use a basis of smooth functions to represent probability densities, so that the forecast model becomes a linear map in this basis.  Numerically, this linear map is estimated by exploiting a rigorous connection between the discrete time shift map and semi-group solution associated to the backward Kolmogorov equation. In \cite{bgh:15}, it was shown that the resulting model estimates the semigroup solutions of the generator of the underlying dynamics in the limit of large data. Moreover, the smooth basis is defined on the training data set, using the diffusion maps algorithm \cite{cl:06,bh:15vb}, which means that the data requirements only depend on the intrinsic dimensionality of the dynamics. 

In this paper, we test this nonparametric modeling approach as a method of forecasting noisy observations of Fourier modes from a selection of well-studied high-dimensional dynamical systems in various turbulent regimes. A novel aspect of this paper is that we consider building a forecasting model given a noisy training data set consisting of partial observations of the dynamics, as is common in practical applications, in contrast to the work in \cite{bgh:15} which used noiseless full observations to train the diffusion forecasting model. A key ingredient for solving initial value problems in any forecasting problem is accurate initial conditions. While initial conditions were assumed to be given in \cite{bgh:15}, in this paper, we introduce a novel Bayesian filtering method to iteratively assimilate each observation and find the initial probability densities given all of the past noisy observations up to the corresponding initial time.

We should note that the diffusion forecasting method \cite{bgh:15} could be naively applied to signals corrupted by observation noise, however the resulting nonparametric model would implicitly include the observation noise in the model, which would limit the forecast skill compared to treating the noise as a separate process.  Treating the noise as a separate process requires first learning the `correct' model from the noisy training data set, and then generating `clean' initial conditions for forecasting from the noisy observations.  In \cite{DMDC,giannakisMajda,GiannakisPNAS} it was shown that applying diffusion maps to the delay-embedded data reduces the influence of the noise on the diffusion maps basis. Building upon the work in \cite{bgh:15}, we apply the theory of \cite{DMDC} to show that building the nonparametric model using the delay-embedded data biases the geometry of the data in a way which extracts the most predictable component of the dynamics. We extend the theory of \cite{DMDC} by giving a rigorous justification for the reduction of the influence of independent additive observation noise on the resulting diffusion forecast model. 


One interesting question which we address here is whether it is possible to build a skillful nonparametric forecasting model for a turbulent mode given only a small amount of noisy training data, when the true dynamics are solutions of a high-dimensional dynamical system with chaotic behavior. This question arises because the nonparametric model has a practical limitation in terms of modeling dynamics with high-dimensional attractors, namely: it will require an immense amount of data to unwind the attractors since the required data increases exponentially as a function of the dimension of the attractor. Moreover, even given a sufficiently large data set, the required computational power would be a limiting factor since the diffusion maps algorithm requires storing and computing eigenvectors of a sparse $N\times N$ matrix, where $N$ is the number of data points.  Constrained by a small data set, the curse-of-dimensionality implies that we cannot unwind the full high-dimensional attractor. We attempt to circumvent the curse-of-dimensionality by decomposing the data into Fourier modes in the hope that delay reconstruction of each mode projects onto a different component of the dynamics. We do not claim that the Fourier decomposition can completely resolve this issue but we will numerically demonstrate that the Fourier decomposition will map an isotropic turbulent field in the spatial domain (which implies that each spatial component is as predictable as any other spatial component) to a coordinate system in which some modes are more predictable than others. Of course, the standard theory of embedology \cite{SYC} suggests that the delay-embedding of a single Fourier mode would reconstruct the entire high-dimensional attractor, which would again be inaccessible to our nonparametric model due to the curse-of-dimensionality. This would suggest that nothing could be gained by building separate models based on delay-embedding of each mode. However, the full attractors reconstructed from each mode are only equivalent in a topological sense, and the geometries of these reconstructed attractors are dramatically different. The biased geometry influences the nonparametric model of \cite{bgh:15} through the use of the diffusion map algorithm which is known to preserve the geometry which the data inherits from the embedding space \cite{cl:06,DMDC}. The diffusion maps algorithm preserves the biased geometry of the delay embedding as was shown in \cite{DMDC}; and we will see that this biased geometry projects the full dynamical system onto the most stable components of the dynamics in the direction of the chosen observations.  When we apply the nonparametric model of \cite{bgh:15} using the basis arising from this biased geometry, we find improved forecasting skill and robustness to observation noise.

The remainder of this paper is organized as follows. In Section~2, we introduce the problems under consideration and establish the necessary background, including a brief overview of the nonparametric modeling approach introduced in \cite{bgh:15} as well as a discussion on how the theory of \cite{DMDC} is applied to mitigate the effect of noise on the model. We conclude Section~2 by introducing the iterative Bayesian filter which we use to generate initial conditions for forecasting with the nonparametric model. In Section~3, we numerically compare predicting Fourier modes of the Lorenz-96 model in various chaotic regimes using the nonparametric model with the persistence model, perfect model, and various parametric models, including the autoregressive models of order-1 (MSM \cite{mh:12}). In Section~4, we numerically compare the nonparametric model with a stochastic model with additive and multiplicative noises (SPEKF model \cite{ghm:10a,ghm:10b}) in predicting the barotropic modes of a geostrophic turbulence. We close this paper with a short summary in Section~5.

\section{Nonparametric diffusion modeling}\label{method}
Let $u(x,t)\in\mathbb{R}^s$ be solutions of an ergodic system of nonlinear PDEs,
\BEA
\frac{\partial u}{\partial t} = \mathcal{A}(u),\label{pde}
\EEA 
where $\mathcal{A}$ denotes nonlinear differential operators, for smooth initial conditions $u(x,0)$ and periodic boundary conditions on a non-dimensionalized periodic domain $x\in[0,2\pi]^n$. To simplify the exposition, set $s,n=1$ without loss of generality. Here, the solutions of \eqref{pde} can be described by the infinite Fourier series,
\BEA 
u(x,t) = \sum_{k\in\mathbb{Z}} \hat{u}_k e^{\mathi kx}, \quad \hat{u}_k^* = \hat{u}_{-k}, \quad \hat{u}_0\in \mathbb{R} \nonumber
\EEA
where the Fourier modes $\hat{u}_k$ can be utilized in analyzing \eqref{pde}. 

Define $\vec{u}_i= (u(x_0,t_i),\ldots,u(x_{2m},t_i)) \in\mathbb{R}^{2m+1}$ whose components are the solutions of \eqref{pde} at time $t_i$ realized at grid point $x_\ell=\ell h, \ell=0,\ldots, 2m$, such that $(2m+1)h = 2\pi$. 
Our goal is to construct a forecasting model for the discrete Fourier coefficients, 
\BEA
\hat{u}_{i,k} =  (\vec{u}_i, \vec{e}^{\,\ell})_h \equiv \frac{h}{2\pi} \sum_{\ell=0}^{2m} u(x_\ell,t_i) e^{-\mathi kx_\ell}, \label{FT}
\EEA
given the corresponding observed modes, $\hat{v}_{i,k} = (\vec{v}_i,\vec{e}^{\,k})_h$ 
where the $\ell$-th component of $\vec{v}_i$, 
\BEA
v(x_\ell,t_i) = u(x_\ell,t_i) + \varepsilon_{\ell,i}, \quad\varepsilon_{\ell,i}\sim\mathcal{N}(0,R),\label{obsmodel} 
\EEA
is the solution of \eqref{pde}, corrupted by i.i.d.~Gaussian noise, at the grid point $x_\ell$ and time step $t_i$. Notice that $\{\vec{e}^{\,\ell}\}_{\ell=0,\pm 1,\ldots,\pm m}$ forms an orthonormal basis of $\mathbb{C}^{2m+1}$ with respect to the inner product defined in \eqref{FT}. One should also note that $\hat{v}_{i,k} = \hat{u}_{i,k} + \eta_{i,k},$ where $\eta_{i,k}\sim\mathcal{N}(0,\hat{R})$ and $\hat{R}=R/(2m+1)$ \cite{mg:09,mh:12}. 

Given the noisy time series of modes $\hat{v}_i := \{\hat{v}_{i,k}\}_{k\in{\cal K}, \{i=1,\ldots,N\}}$, our goal is to use this noisy data set to train the nonparametric model and to generate initial conditions for forecasting modes $\hat{u}_i :=\{\hat{u}_{i,k}\}_{k\in{\cal K}}$ at time index $i>N$. Particularly, we will consider ${\cal K}$ to be a set containing a single mode or containing three modes in our numerical experiments. In Section~\ref{sec21}, we provide an overview of the construction of the nonparametric model introduced in \cite{bgh:15} for fully observed data (all modes) without observation noise.  In Section~\ref{sec22}, we show that by applying a delay embedding on noisy data, $\hat{v}_{i,k}$, we can improve the forecasting skill of the underlying mode, $\hat{u}_{i,k}$. Finally, in Section~\ref{sec23}, we introduce a simple Bayesian filtering method for generating initial conditions for forecasting $\hat{u}_i$ from the noisy observations $\hat v_{i}$.

\subsection{Nonparametric forecasting model}\label{sec21}

In this section, we review the nonparametric diffusion forecasting model introduced in \cite{bgh:15}, assuming the data set consists of full observations of the dynamical system with no observation noise. Consider a system of SDEs,
\BEA
d\hat{u} = a(\hat{u})\,dt + b(\hat{u}) \, dW_t,\quad \hat{u}(0) = \hat{u}_0,\label{SDE}
\EEA
where $a(\hat{u})$ is a vector field, $b(\hat{u})$ is a diffusion tensor, and $W_t$ is an i.i.d.~Wiener process; all defined on a manifold $\mathcal{M}\subset\mathbb{R}^n$. Assume that the dynamical system governed by \eqref{SDE} is ergodic on $\mathcal{M}$ with an invariant measure with density function $\peq(\hat u)$.  We also assume that the evolution of a density function is given by $p(\hat{u},t)=e^{t\mathcal{L}^*}p(\hat{u},0)$ which converges to the invariant measure $\peq(\hat{u})$, where $\mathcal{L^*}$ denotes the Fokker-Planck (or forward) operator and $p(\hat u,0)$ denotes an initial density. The approach we will describe below is nonparametric in the sense that it does not assume any parametric structure on the dynamics, $a$, $b$, the distribution, $p$, or even the manifold, $\mathcal{M}$, which will all be implicitly represented in the model. However, this does not mean that the \emph{method} does not have parameters, and these parameters are roughly analogous to the bin size in a histogram.

Given a time series $\{\hat{u}_i=\hat{u}(t_i)\}_{i=1}^{N}$, our goal is to approximate $p(\hat{u},t)$ without knowing or even explicitly estimating $a, b$. Instead, we will directly estimate the semigroup solution $e^{t\mathcal{L}}$ associated to the generator $\mathcal{L}$ of \eqref{SDE} by projecting the density onto an appropriate basis for $L^2(\mathcal{M},\peq)$. In particular, we choose eigenfunctions $\{\varphi_j\}$ of the generator $\hat{\mathcal{L}}$ of a stochastically forced gradient system,
\BEA
d\tilde{u} = -\nabla U(\tilde{u})\,dt + \sqrt{2}dW_t,\label{gradflow}
\EEA
where the potential function $U \equiv -\log(\peq)$ is defined by the invariant measure of the full system \eqref{SDE} so that the invariant measure of \eqref{gradflow} is $\hat{p}_{eq} = e^{-U} = \peq$. This choice of basis is motivated by several considerations. First, we can estimate the eigenfunctions $\{\varphi_j\}$ by the diffusion maps algorithm for data lying on a compact manifold \cite{cl:06} and on a non-compact manifold \cite{bh:15vb}. In this paper, we will use the variable bandwidth kernel introduced in \cite{bh:15vb} to construct these eigenfunctions since the sampling measure of the data set may be arbitrarily small, which would imply that the data does not lie on a compact manifold. Second, by projecting the density function on this
basis of eigenfunctions, we can accurately forecast each projected component by a discrete representation of the semigroup solution $e^{\tau\mathcal{L}}$, where $\tau=t_{i+1}-t_i$.  We now show how to construct the discrete representation of $e^{\tau\mathcal{L}}$ using the ``shift operator".

Let $\{\varphi_j\}$ be the eigenfunctions of the generator $\hat{\mathcal{L}}$ of the gradient system in \eqref{gradflow}; these eigenfunctions are orthonormal under $\langle\cdot,\cdot\rangle_\peq$ in $L^2(\mathcal{M},\peq)$ and the integral here (and in all of the inner product defined below) is with respect to the volume form $dV$ which $\mathcal{M}$ inherits from the ambient space. Note that $\hat{\mathcal{L}}$ is the generator of gradient system in \eqref{gradflow} and it is the adjoint of the Fokker-Planck operator $\hat{\mathcal{L}}^*$ with respect to inner-product $\langle\cdot,\cdot\rangle$ in $L^2(\mathcal{M})$.  One can show that $\{\psi_j = \varphi_j\peq\}$ are eigenfunctions of $\hat{\mathcal{L}}^*$ which are orthonormal with respect to inner-product $\langle\cdot,\cdot\rangle_{\peq^{-1}}$ in $L^2(\mathcal{M},\peq^{-1})$. Given an initial density $p(\hat{u},0)=p_0(\hat{u})$, we can write the forecast density $p(\hat{u},t)$ as,
\BEA
p(\hat{u},t) = e^{t\mathcal{L}^*}p_0(\hat{u}) = \sum_j \langle e^{t\mathcal{L}^*}p_0,\psi_j \rangle_{\peq^{-1}}\psi_j(\hat{u}) =  \sum_j \langle p_0,e^{t\mathcal{L}}\varphi_j \rangle\varphi_j(\hat{u})\peq(\hat{u}).\label{pexpand}
\EEA
Setting $t=0$ in \eqref{pexpand} we find,
\BEA
p_0(\hat{u}) = p(\hat{u},0) =\sum_j \langle p_0,\varphi_j \rangle\varphi_j(\hat{u})\peq(\hat{u})  \equiv \sum_j c_j(0)\varphi_j(\hat{u})\peq(\hat{u}),\label{po}
\EEA
where we define $c_j(0) \equiv \langle p_0,\varphi_j \rangle$. Substituting \eqref{po} into \eqref{pexpand}, we obtain,
\BEA
p(\hat{u},t) = \sum_j\sum_l c_l(0)\langle \varphi_l,e^{t\mathcal{L}}\varphi_j \rangle_\peq\varphi_j(\hat{u})\peq(\hat{u}).\label{p}
\EEA
The key idea of the non-parametric forecasting algorithm in \cite{bgh:15} is to approximate $A_{jl}\equiv\langle\varphi_l,e^{\tau\mathcal{L}}\varphi_j \rangle_\peq$ by replacing the semi-group solution $e^{\tau\mathcal{L}}$ with the discrete time shift operator $S$, which is defined as $Sf(\hat{u}_i) = f(\hat{u}_{i+1})$ for any $f\in L^2(\mathcal{M},\peq)$.  In \cite{bgh:15}, it was shown that $S$ is a stochastic estimate of $e^{\tau\mathcal{L}}$ and the error due to the stochastic terms can be minimized by projecting $S$ on the basis $\varphi_j$. Indeed it was shown that $\hat A_{jl} \equiv \langle\varphi_l,S\varphi_j \rangle_\peq$ is an unbiased estimate of $A_{jl}$, meaning that $\mathbb{E}[\hat A_{jl}] = A_{jl}$. Furthermore, assuming that $\hat{u}_i$ are independent samples of $\peq$, one can show that the error of this estimate is of order $\sqrt{\tau/N}$, which means that one can apply this approximation for any sampling time $\tau$ given a sufficiently large data set $N$. 

Notice that for any $f\in\mathcal{L}^2({\cal M},\peq)$, we have:
\BEA
\langle \varphi_l,e^{\tau\cal L}f \rangle_\peq &=& \langle \varphi_l,e^{\tau\cal L} \sum_j \langle f,\varphi_j\rangle_\peq \varphi_j \rangle_\peq \nonumber \\ &=& \sum_j \langle \varphi_l,e^{\tau\cal L}\varphi_j \rangle_\peq \langle f,\varphi_j \rangle_\peq. \label{AA}
\EEA
From the ergodicity property of \eqref{SDE}, one can deduce that the largest eigenvalue of $e^{\tau\mathcal{L}}$ is equal to 1 with constant eigenfunction, $\mathbbm{1}(\hat u)$. Setting $f= \mathbbm{1}$ in \eqref{AA}, we have, 
\BEA
\sum_j \langle \varphi_l,e^{\tau\cal L}\varphi_j \rangle_\peq \langle \mathbbm{1},\varphi_j \rangle_\peq = \langle \varphi_l,e^{\tau\cal L}\mathbbm{1}\rangle_\peq  = \langle \varphi_l, \mathbbm{1}\rangle_\peq,\nonumber
\EEA
which implies that $\sum_j A_{lj} \langle \mathbbm{1},\varphi_j \rangle_{\peq}= \langle \mathbbm{1},\varphi_l \rangle_\peq$ or in compact form, $A\vec{e}_1 = \vec{e}_1$, where $[\vec{e}_1]_j=\langle \mathbbm{1}, \varphi_j\rangle_\peq$, so $\vec{e}_1$ is $1$ on the first component and zero otherwise. Assuming that the data is sampled from $\peq$, we just deduced that the largest eigenvalue of $A$ should be equal to 1 (with no complex component). Numerically however, we sometimes find that $\hat{A}$ has some eigenvalues slightly larger than one, due to the finite number of samples in the Monte-Carlo integrals. If this occurs, we can easily ensure the stability of the diffusion forecast by dividing any eigenvalue with norm greater than 1 so that is has norm equal to one. We note that in our numerical experiments below this issue rarely occurs.  The ease with which we can guarantee the stability of this nonparametric model is an advantage over many parametric methods.

We should also note that if \eqref{SDE} is a gradient system as in \eqref{gradflow}, then $A_{jl}\equiv\langle\varphi_l,e^{\tau\mathcal{L}}\varphi_j \rangle_\peq = e^{\lambda_j\tau}\delta_{j,l}$, where $\lambda_j$ are the eigenvalues of $\mathcal{L}=\hat{\mathcal{L}}$, which can be obtained directly from the diffusion maps algorithm \cite{cl:06,bh:15vb}. Moreover, the matrix $A$ becomes diagonal (due to the orthonormality of $\varphi_j$) so that the diffusion maps algorithm estimates $A$ directly and the shift operator approximation is not required. See \cite{bh:14uq} for various uncertainty quantification applications for this special case.

To conclude, the non-parametric forecasting algorithm (which we refer to as the {\it diffusion forecast}) for a given initial density $p_0(\hat{u})$ is performed as follows:
\begin{enumerate}
\item {\it Learning phase}: Given a data set $\{\hat{u}_i\}_{i=1}^N$, apply the diffusion maps algorithm \cite{cl:06,bh:15vb} to obtain eigenvectors $\{\vec{\varphi}_j\}$, whose $i$-th component, $(\vec{\varphi}_j)_i =\varphi_j(\hat{u}_i)$, approximates the eigenfunction $\varphi_j$ of the gradient flow \eqref{gradflow} evaluated at the data point $\hat{u}_i$. We implement the diffusion maps algorithm with a variable bandwidth kernel, see the Appendix for the step-by-step algorithm to obtain these eigenfunctions. Also, see the supplementary material of \cite{bgh:15} for a short overview of \cite{bh:15vb}.
\item {\it Initial conditions}: Represent the initial density $p_0$ in this basis as,
\BEA
c_j(0) &=& \langle p_0,\varphi_j \rangle = \langle p_0/\peq,\varphi_j \rangle_\peq =  \sum_{i=1}^N \frac{p_0(\hat{u}_i)}{\peq(\hat{u}_i)}\varphi_j(\hat{u}_i). 
\EEA
which is numerically obtainable from the last inner product as a Monte-Carlo integral since we have $\varphi_j$ evaluated on 
the data set $\hat{u}_i$ which are samples of $\peq$. 
\item {\it Forecasting}: Apply a Monte-Carlo integral to approximate the shift operator $S$ in coordinate basis $\varphi_j$,
\BEA
\hat{A}_{jl}\equiv  \langle\varphi_l,S\varphi_j \rangle_\peq \approx \frac{1}{N}\sum_{i=1}^N \varphi_l(\hat u_i)\varphi_j(\hat u_{i+1}).
\EEA
We then approximate $A_{jl}$ with $\hat{A}_{jl}$ such that the diffusion forecast is given by
\BEA
p(\hat u,m\tau) \approx \sum_j\sum_l (\hat{A}^m)_{jl} c_l(0)\varphi_j(\hat u)\peq(\hat u)\equiv\sum_{j}c_j(m\tau)\varphi_j(\hat u)\peq(\hat u),
\EEA
where in practice, these sums are truncated at a finite number, $M$, of eigenfunctions.  With this truncation, the coefficient $c_j(m\tau)$ is the $j$-th component of a matrix-vector multiplication, $\vec{c}(m\tau)=\hat{A}^m\vec{c}(0)$, of an $M\times M$ matrix $\hat{A}^m$ and an $M$-dimensional vector $\vec{c}(0)=(c_1(0),\ldots,c_M(0))^\top$.  
\end{enumerate}

 \subsection{Time-lagged embedding}\label{sec22}

The diffusion forecast algorithm of Section~\ref{sec21} assumes that the data $\hat u_i$ are sampled directly from the full dynamical system \eqref{SDE}.  That is, the full system $\hat u$ in \eqref{SDE} is equivalent to the dynamical system for $\vec u$ and the turbulent dynamics considered here will be very high dimensional.  For such high dimensional problems, exploring the entire attractor would require a prohibitively large data set; larger than will typically be available in applications.  Instead, we attempt to build a low-dimensional model for each mode $\hat u_{i,k}$ individually.  An individual mode $\hat u_{i,k}$ is simply a linear projection of the full system $\hat u_i$, and moreover it is an `observation' of the spatial state $u$.  While we could apply the diffusion forecast to the one dimensional time series $\hat u_{i,k}$, this would ignore the dependence of the $k$-th mode on all the other modes.  The fundamental problem with building a one-dimensional stochastic model for $\hat u_{i,k}$ is that any interaction with the other modes will result in (among other changes to the model) an inflated stochastic component in the closed model for $\hat u_{i,k}$ (see \cite{gh:13,PRSA} for a rigorous example).  Inflating the stochastic component of the model for $\hat u_{i,k}$ will severely limit the predictive skill of the nonparametric model. On the other hand, if we include other modes in the nonparametric model, this would increase the dimensionality and not all of the information in the other modes would be relevant to forecasting the $k$-th mode.  Instead, we will apply the delay coordinate reconstruction of \cite{Takens,SYC,stark1,stark2} to implicitly recover only the missing components of the dynamics which are important for forecasting each $\hat u_{i,k}$.  Moreover, we will apply the theory of \cite{DMDC} to show that the delay coordinate reconstruction projects the missing components of the dynamics onto the most stable component of $\hat u_{i,k}$.  Finally, in practice we will only have access to a noisy data set $\hat v_{i,k}$, and the theory of \cite{DMDC} suggests that the delay coordinate reconstruction also reduces the influence of the observations noise, which we will now make rigorous.

Given a time series $\hat u_{i,k} = \hat u_k(t_i)$ of the $k$-th mode, we construct the delay embedding coordinates,
\[ \vec{\hat{u}}_{i,k} = H(\hat u_{i,k}) = (\hat u_{i,k},\hat u_{i-1,k},...,\hat u_{i-L,k})^\top. \]
The theory of embedology shows that if the $k$-th mode is a generic observation of the full system \eqref{SDE}, then for sufficiently many lags, $L$, there exists a diffeomorphism $\mathcal{F}$ such that $\vec{\hat{u}}_{i,k} = \mathcal{F}(\hat u_i)$.  This statement was first shown for deterministic dynamics on smooth attractors in \cite{Takens} and subsequently generalized to fractal attractors in \cite{SYC} and then to non-autonomous systems in \cite{stark1} and stochastic systems in \cite{stark2}.  The existence of the diffeomorphism $\mathcal{F}$ says that topologically the attractor of the delay reconstruction $\vec{\hat{u}}_{i,k}$ is equivalent to the full dynamics $\hat u_i$ and so we have \emph{reconstructed} the hidden variables in the evolution of the $k$-th mode.  However, the theory of embedology only concerns the topology of the attractor, whereas the basis of eigenfunctions $\{\varphi_j\}$ that are estimated from the diffusion maps algorithm will depend on the geometry of the delay reconstruction $\vec{\hat{u}}_{i,k}$.  

In \cite{DMDC} it was shown that the delay reconstruction severely biases the geometry in the reconstructed coordinates $\vec{\hat{u}}_{i,k}$, so that in the limit of infinite delays, $L\to \infty$, the dynamics are projected on the most stable component of the dynamics.  Consider the $k$-th mode $\hat u_{i,k}$ as an observation of the state $\hat u_i$, where the observation function is given by,
\[ \hat u_{i,k} = h_k(\hat u_i) = (0,\ldots,0,1,0,\ldots,0) \hat u_i. \]  
Notice that the derivative of the observation function is $Dh_k = (0,\ldots,0,1,0,\ldots,0)$ where the $1$ occurs in the $k$-th entry.  Moreover, the previous value of the $k$-th mode, $\hat u_{i-1,k}$, can also be considered an observation of $\hat u_i$ where the observation function is given by $\hat u_{i-1,k} = h_k(F_{-\tau}(\hat u_i))$ and the map $F_{-\tau}(\hat u_i) = \hat u_{i-1}$ is given by the reverse time evolution for the discrete time step $\tau$.  Interpreting each $\hat u_{i-l,k}$ as an observation of the full state $\hat u$ at time $t_i$ shows that the time-delay embedding $\vec{\hat u}_{i,k}$ is an observation of $\hat u_i$ with observation function $H$ as follows,
\BEA
\vec{\hat{u}}_{i,k} = H(\hat u_{i,k})  = \Big(h_k(\hat{u}_{i}),h_k(F_{-\tau}(\hat{u}_{i})),\ldots,h_k(F_{-L\tau}(\hat{u}_{i}))\Big)^\top,\nonumber
\EEA 
where $F_{-l\tau}$ is the reversed shift map which takes $\hat u_{i}$ to $\hat u_{i-l}$.
For $L$ sufficiently large and assuming that the $k$-th mode is a generic observable on the manifold, the observation $H$ will have a full rank derivative \cite{SYC}.  Our goal now is to examine the change in metric induced by this derivative by seeing how it acts on tangent vectors on the attractor.  Let $\nu_1, \nu_2 \in T_{\hat u}\mathcal{M}$ be tangent vectors on the attractor and let $\hat{\nu_1}=DH(\hat u)\nu_1$ and $\hat{\nu_2}=DH(\hat u)\nu_2$ where $\hat{\nu}_1,\hat{\nu}_2 \in T_{H(\hat u)}H(\mathcal{M})$ are the transformed tangent vectors in the new geometry given by the time-delay embedding. Then the inner product $\langle \cdot, \cdot \rangle_{\mathbb{R}^L}$ in the delay embedding space is
\begin{align}
\langle \hat{\nu_1},\hat{\nu_2} \rangle_{\mathbb{R}^L} &= \sum_{l=0}^L \langle Dh_k(F_{-l\tau}(\hat u))DF_{-l\tau}(\hat u)\nu_1,
Dh_k(F_{-l\tau}(\hat u))DF_{-l\tau}(\hat u)\nu_2 \rangle_{\mathbb{R}} \nonumber \\
&= \sum_{l=0}^L (DF_{-l\tau}(\hat u)\nu_1)_k (DF_{-l\tau}(\hat u)\nu_2)_k ,
\end{align}
where we used the fact that $Dh_k = (0,\ldots,0,1,0,\ldots,0)$ and the subscript-$k$ denotes the $k$th component of the corresponding vector.  If $\nu_1$ and $\nu_2$ are in the $m$-th Oseledets space, with Lyapunov exponent $\sigma_m$, then the inner product reduces to, $\langle \hat{\nu_1},\hat{\nu_2} \rangle_{\mathbb{R}^L} =\sum_{l=0}^L e^{-2\sigma_m l}(\nu_1)_k(\nu_2)_k$.  This shows that the most stable Oseledets space will dominate the geometry in the embedding since $\sigma_m<0$ will be most negative (largest in absolute value of the negative Lyapunov exponents) in the most stable direction.  

The bias introduced into the geometry by the delay embedding has several advantages for forecasting.  First, the most stable components of the dynamics are the easiest to predict since trajectories converge very quickly when measured in these components.  Given an initial condition which is a small perturbation of the truth in a stable direction, the perturbation will decrease in the forecast, reducing the effect of the initial perturbation.  Since numerically we will not be able to represent the whole attractor, the variable bandwidth kernel will implicitly project away the small components of the geometry.  By amplifying the stable components we insure that the most desirable aspects of the dynamics will be well represented in the discrete representation of the geometry.  Secondly, when applied to the noisy data, $\hat v$, the noise will correspond to an unstable component, and so the delay embedding geometry will de-emphasize the noise component of the data.  Formally, consider a noisy observation $\hat v_{i,k} = \hat u_{i,k} + \xi_i$ where $\xi_i$ are independent identically distributed random variables independent of $\hat u_{i,k}$ with $\mathbb{E}[\xi_i]=0$.  When we compute the inner product of noisy vectors in the delay embedding space we find the relative difference, 
\[  H(\hat v_{i,k})^\top H(\hat v_{j,k}) = H(\hat u_{i,k})^\top  H(\hat u_{j,k}) + \sum_{l=0}^L \xi_{i-l}\hat u_{j-l,k} + \xi_{j-l}\hat u_{i-l,k} + \xi_{i-l} \xi_{j-l}. \]
Since the noise $\xi_i$ is independent and independent of $\hat u_{\cdot,k}$ all the terms inside the sum above have expected value of zero, so by the law of large numbers, in the limit as $L\to \infty$ we find that for $i\neq j$,
\begin{align} & \frac{H(\hat v_{i,k})^{\top} H(\hat v_{j,k}) -H(\hat u_{i,k})^\top H(\hat u_{j,k})}{H(\hat u_{i,k})^\top H(\hat u_{j,k})} \nonumber \\ &= \frac{\frac{1}{L}\sum_{l=0}^L \xi_{i-l}\hat u_{j-l,k} + \xi_{j-l}\hat u_{i-l,k} + \xi_{i-l} \xi_{j-l}}{\frac{1}{L}H(\hat u_{i,k})^\top H(\hat u_{j,k})}\rightarrow 0, \end{align}
assuming that ${\frac{1}{L}H(\hat u_{i,k})^\top H(\hat u_{j,k})}=\frac{1}{L}\sum_{l=0}^L \hat{u}_{i-l,k}\hat{u}_{j-l,k}$ is nonzero as $L\rightarrow\infty$. Since we preserve all the pairwise inner product in the embedding space, this implies that we preserve the metric up to a constant. This shows how the additive observation noise has a smaller impact on the delay embedding than the original embedding, especially for large $L$. 

 Finally, note that the delay reconstruction of each mode, $\vec{\hat u}_{\cdot,k}$, will capture a different aspect of the most stable components of the dynamics since $(\nu_1)_k(\nu_2)_k$ corresponds to the projection of the vectors $\nu_1,\nu_2$ into the $k$-th coordinate direction.  Thus, the delay embedded modes $\vec{\hat v}_{i,k} = (\hat v_{i,k},...,\hat v_{i-L,k})^\top$ represent orthogonal components of the most stable Oseledets directions of the dynamics, which in turn are orthogonal to the additive observation noise in the limit of large $L$.

\begin{figure}
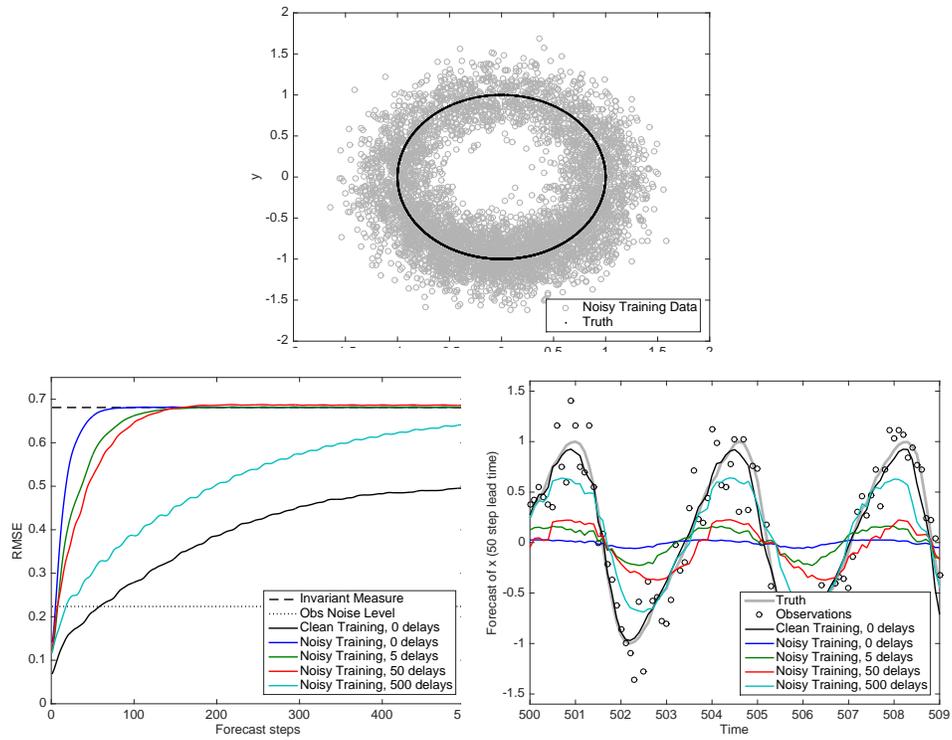

\begin{center}
\includegraphics[width=.45\linewidth]{CircleTrainingData.eps} \\
\includegraphics[width=.45\linewidth]{CircleForecastRMSE.eps}
\includegraphics[width=.45\linewidth]{CircleForecast.eps}
\end{center}
\caption{For training data with a significant amount of additive observation noise (top), the effect of increasing the number of lags on the diffusion model learned from noisy observations is shown in terms of the forecast RMSE (bottom, left) and the 50 step lead time forecast (bottom, right).}
\label{circleEx}
\end{figure}

We now demonstrate the effect of the biased geometry on a simple example.  We will generate a stochastic dynamics on the unit circle by first numerically integrating the one-dimensional SDE,
\begin{align}\label{circleSDE} d\theta = (2+\sin(\theta))\, dt + \sqrt{0.1} \, dW_t, \end{align}
and then mapping the intrinsic variable $\theta$ onto the unit circle embedded in $\mathbb{R}^2$ by the map $\theta \mapsto (x(\theta),y(\theta))^\top = (\cos(\theta),\sin(\theta))^\top$. In this example we use discrete time step $\Delta t = 0.1$ to produce 10000 samples $\theta_i$ of the system \eqref{circleSDE} which are mapped into the plane as $(x_i,y_i)^\top = (\cos(\theta_i),\sin(\theta_i))^\top$.  We then generate noisy observations $(\tilde x_i,\tilde y_i)^\top = (x_i,y_i)^\top + \eta_i$ where the observation noise $\eta_i$ are independently sampled from a two dimensional Gaussian distribution with mean zero and covariance matrix $\frac{1}{20} I_{2\times 2}$.  We chose this example to be very predictable given the cleaning training data $(x_i,y_i)^\top$, and the small stochastic component is only included so that even the perfect model can only predict for a finite time.  Notice that the observation noise in this example is very large, as shown in Figure \ref{circleEx}, for example the variance of the noise in the x-coordinate is approximately 32\% of the variance of the true signal.  

Our goal in this example is to show how the delay-embedding can reduce the influence of observation noise on a diffusion model by biasing the geometry.  We use the first 5000 noisy data points $(\tilde x_i,\tilde y_i)^\top$ to train multiple diffusion models, each time applying a different number of delays before building the diffusion model.  We also train a diffusion model using the clean data set $(x_i,y_i)^\top$.  For each diffusion model, we apply the Bayesian filter developed in Section~\ref{sec23} below to generate initial conditions for forecasting from the noisy observations $(\tilde x_i,\tilde y_i)^\top$ in the verification period $i=5001,...,10000$.  Each initial condition in the verification period is forecast for a total of 500 forecast steps (50 time units) and the RMSE between the forecast and the truth is averaged over the verification period.  The RMSE as a function of the number of forecast steps for each model is shown in Figure \ref{circleEx}, along with the forecast of the $x$ coordinate at the 50 step lead time, compared to the true value of $x$.   

While no amount of delays is able to match the diffusion model built using the clean data set, the improvement in forecast skill is significant as the number of delays increases.  We should note that since this model is intrinsically one dimensional, the most stable component of the dynamics represents the entire dynamical system.  This implies that projecting on the most stable component can only improve the forecasting skill for this example.  For high-dimensional examples, as the number of delays becomes very high, the projection onto the most stable component will remove information which is important to the forecast, leading to reduced forecast skill as shown in Figure \ref{L96_varyL} which we will discuss in Section 3 below.  Therefore, in general we expect to see a tradeoff where small numbers of delays can remove the influence of noise, but extremely large numbers of delays may actually degrade the forecast by projecting away valuable components of the dynamics.

\subsection{Bayesian filter}\label{sec23}

In Step~2 of the forecasting algorithm described in Section~\ref{sec21}, we assume that the initial distributions for forecasting are given. In practice, however, we only have the noisy data $\hat{v}_i:=\{\hat{v}_{i,k}\}_{k \in \cal{K}}$, and the goal of this section is to develop a numerical method to determine the initial distribution of the corresponding modes, $\hat{u}_i:=\{\hat{u}_{i,k}\}_{k \in \cal{K}}$.  We will introduce a Bayesian filter to iteratively combine the information from the noisy data with the diffusion forecast. We should note that the noisy observations which we use for forecasting are always taken out-of-sample, meaning they are separate from the training data. 

If the observations were continuous (i.e. for $\tau\rightarrow 0$), we could filter the noisy observations $\hat v_i$ using the Zakai equation projected onto the basis $\{\varphi_j\}$, and this was the approach taken in \cite{bh:14uq}.  However, since the observation time may be long, we will take a different approach here and apply a Bayesian update at discrete finite time step.  Explicitly, we want to find the posterior density $p(\hat u_i \, | \, \hat v_{s\leq i})$ by updating the previous posterior density $p(\hat u_{i-1} \, | \, \hat v_{s\leq i-1})$ with the current noisy observation $\hat v_i$.  

Our Bayesian filter will follow the standard predictor-corrector approach.  Given an initial density at the previous time, $p(\hat u_{i-1} \, | \, \hat v_{s\leq i-1})$, the first step is to compute the prior forecast density with our nonparametric model, 
\[ p(\hat u_i \, | \, \hat v_{s \leq i-1}) = \sum_{j=1}^M c_j(t_i)\varphi_j(\hat u)\peq(\hat u) = \sum_{j=1}^M (\hat A c(t_{i-1}))_j \varphi_j(\hat u)\peq(\hat u), \] 
where the coefficients $c_j(t_{i-1}) = \left<p(\hat u_{i-1} \, | \, \hat v_{s\leq i-1}),\varphi_j \right>$ represent the posterior density at time $t_{i-1}$, projected in diffusion coordinates, $\varphi_j$.  Since the noisy observation $\hat v_i$ has the form $\hat{v}_{i} = \hat{u}_{i} + \eta_i$ where $\eta_i\sim\mathcal{N}(0,\hat{R})$, the likelihood function for the noisy observation is,
\[ p(\hat v_i \, | \, \hat u_i) \propto \exp(-||\hat v_i-\hat u_i||_{\hat R}^2/2). \]
We can now assimilate the observation $\hat v_i$ by using Bayes law to combine the prior forecast and the likelihood function above as follows,
\begin{align}\label{bayes} p(\hat u_i \, | \, \hat v_{s \leq i}) &\propto  p(\hat u_i \, | \hat v_{s \leq i-1})p(\hat v_i \, | \, \hat u_i ).  \end{align}
By computing the product \eqref{bayes} we can find the desired posterior at time $t_i$, up to the normalization factor.  We estimate the normalization factor, $Z$, as a Monte-Carlo integral,
\[ Z = \sum_{l=1}^N \frac{p(\hat u_l \, | \hat v_{s \leq i-1})p(\hat v_i \, | \, \hat u_l )}{\peq(\hat u_l)} \approx \int_{\mathcal{M}} p(\hat u \, | \hat v_{s \leq i-1})p(\hat v_i \, | \, \hat u )\, dV(\hat u), \]
and dividing the product in \eqref{bayes} by $Z$ we recover the posterior density at time $t_i$. 

In order to initialize this iterative procedure, we spin up the filtering process for few assimilation cycles starting from the invariant measure $p(\hat u_0) = \peq(\hat{u})$.  To obtain initial density $p(\hat u_{i} \, | \, \hat v_{s\leq i})$, which is used for forecasting starting at time $t_i$, we run this Bayesian filter, starting from $t_{i-P}$ to allow for $P$ steps of spin-up with the invariant distribution as the initial density, $p(\hat u_{i-P} \, | \, \hat v_{s\leq i-P})=\peq(\hat{u})$. In our numerical simulations below, we used a very short spin-up time of  $P=10$ steps.

The method developed above requires only that the observation noise is independent and that the likelihood function function $p(\hat v_i \, | \, \hat u_i)$ is known.  The only assumption on the posterior density is that it is well approximated in the truncated basis $\{\varphi_j\}_{j=1}^M$, which intuitively implies that the posterior density is smooth and is not very highly oscillatory.

\section{Forecasting weakly chaotic to fully turbulent modes of Lorenz-96 model}

As our test example, we consider Fourier modes of the Lorenz-96 (L96) model \cite{lorenz:96} in various chaotic regimes. The Lorenz-96 model is defined by the following forced-dissipative nonlinear system of ODEs,
\BEA
\frac{du_\ell}{dt} = (u_{\ell+1}-u_{\ell-2})u_{\ell-1} - u_\ell +F, \quad \ell=1, \ldots, 40, \label{L96}
\EEA
The right hand side of \eqref{L96} consists of an energy conserving quadratic advective-like term, a linear dissipation, and a forcing parameter $F$. Following \cite{lorenz:96}, we resolve the L96 model at 40 equally spaced grid points with a periodic boundary (represented by the subscript $l$ being taken modulo 40) in order to mimic weather wave patterns on a midlatitude belt. The statistical behavior of the Fourier modes of this model has been analyzed in \cite{mag:05}; as the forcing parameter $F$ increases, the model becomes fully turbulent with Gaussian-like modal distributions. In our numerical experiments we will consider three different values of $F$ following \cite{mag:05}. As reported in \cite{am:07}, for $F=6$, the system is weakly chaotic with largest Lyapunov exponent $\lambda_1 = 1.02$ and the dimension of the expanding subspace of the attractor is $N^+=12$. For $F=8$, which is the original choice in \cite{lorenz:96}, the system is strongly chaotic with $\lambda_1=1.74$ and $N^+=13$. For $F=16$, the system is ``fully turbulent" with $\lambda_1=3.94$ and $N^+=16$. We should also note that when $F=6$, the L96 model has a longer ``memory" in the sense that it has a relatively slow decaying time (which is defined as the integral of autocorrelation function, see e.g., \cite{mag:05}), this is manifested visibly as a regular dominant westward propagating ``weather-like" wave pattern of wavenumber-8. The memory becomes shorter as $F$ increases and the ``weather-like" wave pattern becomes less obvious. 

\subsection{Diagnostic models}

In our numerical experiments we will examine the forecasting skill of the nonparametric modeling approach on a selection of Fourier modes, including those with high and low energies as well as large and small correlation times. To diagnose the forecasting skill, we compare the diffusion forecast on these modes with forecasts from standard statistical approaches when the underlying dynamics are unknown as well as with the perfect model. In particular, we will consider:

\subsubsection{Persistence model}
The simplest nonparametric model for predicting the future when the underlying dynamics are not known is by setting the current observation, $\hat v_i$, as the forecast at each lead time.  

\subsubsection{Linear autoregressive models}
A popular statistical modeling approach when the underlying dynamics are not known is to fit the data to a class of linear autoregressive models. In this paper, we consider two models from such a parametric modeling approach that were designed previously as cheap filter models: the Mean Stochastic Model (MSM) \cite{hm:08b,mh:12} and the stable constrained autoregressive model of order-3 (AR3) \cite{bh:13b,hhr:15}. 

The key idea of MSM is to model each Fourier mode as a complex valued linear stochastic model with additive white noise,
\BEA
d u = \lambda u \,dt + \sigma\,dW_t,\label{msm}
\EEA
where $W_t$ is a complex valued Wiener noise with variance $t$.  The parameters $\lambda$ and $\sigma$ are determined by matching the analytical expression for the variance and correlation time at the statistical equilibrium state of the MSM model in \eqref{msm} with the corresponding empirically estimated statistics from the data (see Chapter~12 of \cite{mh:12} for a detailed formulation). In our implementation below, we will obtain these parameters from the statistics of the noisy dataset, $\hat{v}$, during the training period. To generate initial conditions for the forecast, we apply a one-dimensional Kalman filter in order to account for the observation noise in $\hat{v}$, and we assume that the noise observation variance, $\hat{R}$, is known. 

The stable and consistent AR3 model was introduced in \cite{bh:13b} as a cheap linear model for filtering turbulent modes with long memory, such as the L96 model with $F=6$. The standard AR3 model is given by,
\BEA
u_{m+1} = a_1 u_{m-2} + a_2 u_{m-1} + (1+a_3) u_m + \eta_m, \quad \eta_m\sim\mathcal{N}(0,Q).
\EEA
The stability of this model is sensitive to the choice of sampling time $\Delta t = t_{m+1}-t_m$, especially when a standard linear regression is used to determine the model parameters. In \cite{bh:13b}, it was shown that one can obtain accurate filtered mean estimates in a linear filtering problem with truth generated from \eqref{msm} if the parameters $a_j$ are chosen to satisfy a set of linear algebraic consistency conditions which depends on $\lambda$ of \eqref{msm} and the sampling time $\Delta t$ (see \cite{bh:13b} for details). For nonlinear problems, a stable model together with these consistency constraints can be determined by an algebraic method proposed in \cite{hhr:15} and this AR3 model is what we use in the examples below. In particular, we will enforce the consistency condition for the AR3 model using the parameter $\lambda$ obtained from the MSM fit of the noisy data at the training period.  To generate initial conditions for the forecast, we apply a three-dimensional Kalman filter to account for the noise observed $\hat{v}_k$, assuming that the observation noise variance, $\hat{R}$, is known (see e.g. \cite{kh:12b,bh:13b,hhr:15} for the details of the AR filter).

\subsubsection{Perfect model}
Finally, we also include the forecast of the perfect model in \eqref{L96}. To generate the initial conditions, we implement a standard ensemble Kalman filter method \cite{hunt:07} with 80 ensemble members, double the dimension of the L96 model. The perfect model experiment with full data refers to forecasting skill where the initial conditions are determined by filtering noisy observations at all 40 grid points. 

We also show forecasting skill with a perfect model experiment given only observations of noisy modes $\hat v_i=\{\hat v_{i,k}\}_{k\in{\cal K}}$. In this scenario, the ensemble Kalman filter estimates the initial conditions at unobserved modes which will be needed for integrating the full model; obviously, if every mode is observed, we are back to the perfect model experiment with full data. Comparing this scenario with the diffusion forecast constructed from the same $\hat v_i$ provides the most objective comparison between the full model and the nonparametric diffusion model. In the example below, we will consider examples where ${\cal K}$ consists of a single mode and the three most energetic modes, ${\cal K} =\{7, 8, 9\}$.

\subsection{Experimental design}
In our numerical simulations, the true time series are generated by integrating the L96 model with the fourth-order Runge-Kutta method with time step $\delta t=1/64$. Noisy observations are generated by adding independent random samples of a Gaussian with mean zero and variance $R=1$ to the true solutions at every grid point at each observation time step $\Delta t=1/8$ as in \eqref{obsmodel}, resulting in a noise variance of $\hat{R} = 1/40$ on each Fourier mode. Given a time series of length 15000, we use the first 5000 data points to train the three models (diffusion, MSM, and AR3) and we compute the prediction skill on the remaining $V=10000$ data points. To diagnose the prediction skill, we use the standard Root-Mean-Squared Error (RMSE) and anomaly pattern correlation skill defined as follows,
\BEA
RMSE (\tau) &=& \sqrt{\frac{1}{V}\sum_{i=1}^V(u^t_{i}(\tau)-u^f_i (\tau))^2 },\label{rms}\\
PC(\tau) &=&\frac{1}{V}\sum_{i=1}^V \frac{(u^t_{i}(\tau) - \bar{u}^t)^\top(u^f_{i}(\tau) - \bar{u}^t)}{\| u^t_{i}(\tau) - \bar{u}^t\| \|u^f_{i}(\tau) - \bar{u}^t \|},\label{pc}
\EEA
where $u^t_i(\tau)$ denotes the truth and $u^f_i(\tau)$ denotes the forecast at lead time $\tau$. As defined in \cite{wilks2011}, the anomalies are defined with respect to the climatology which is empirically estimated by time average of the truth, $\bar{u}^t = V^{-1}\sum_{i=1}^V u^t_i$. In our evaluation below, we will evaluate the forecast skill on Fourier modes $k\in{\cal K}$, where $k\neq 0$; the climatological means of these nonzero wave numbers are zero. We will also evaluate these measures on the 40-dimensional spatial domain solutions and in this case, the climatological mean $\bar{u}^t$ is nonzero in our example.

\begin{figure}
\begin{center}
\mbox{
\includegraphics[width=.45\textwidth]{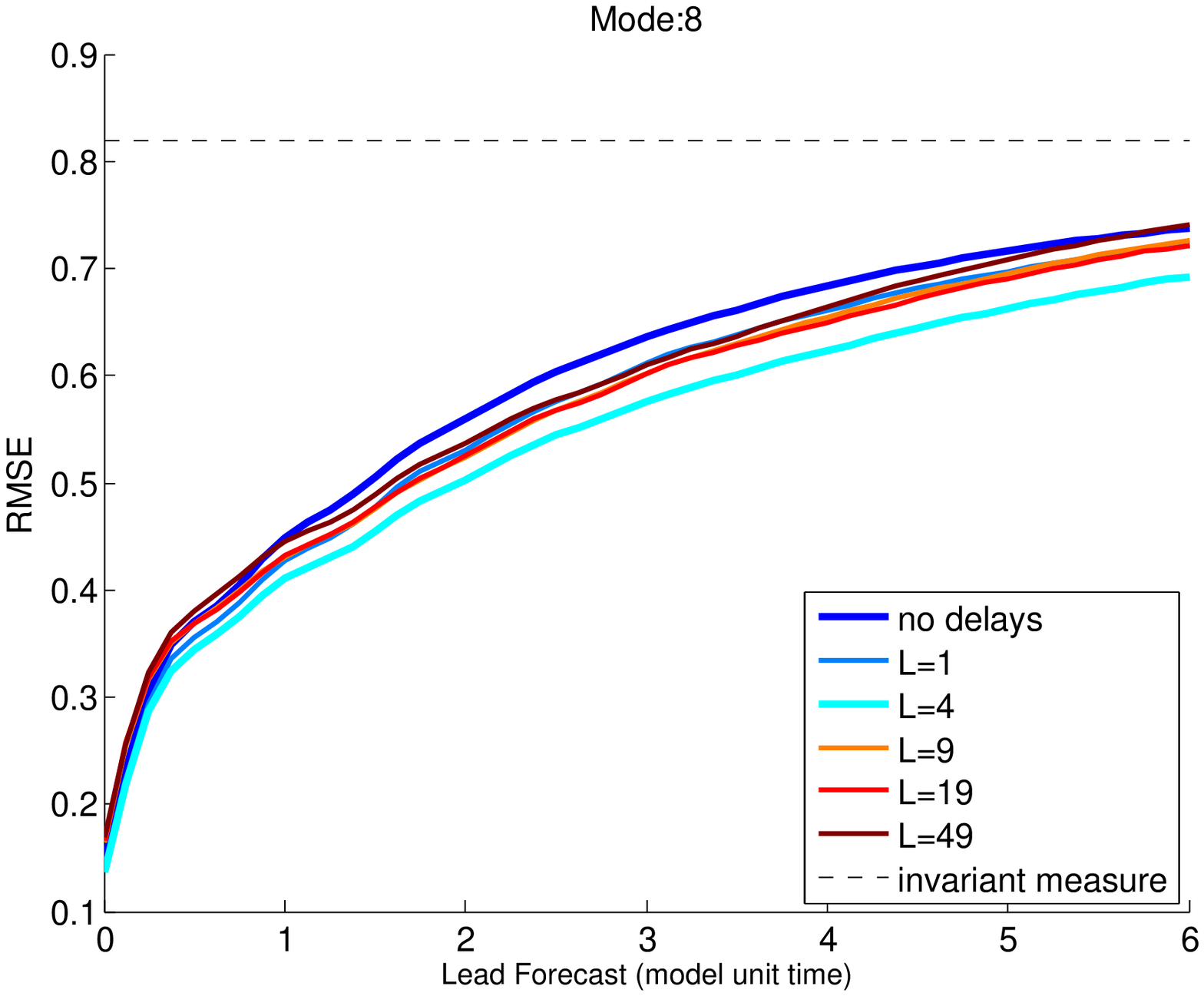}
\includegraphics[width=.45\textwidth]{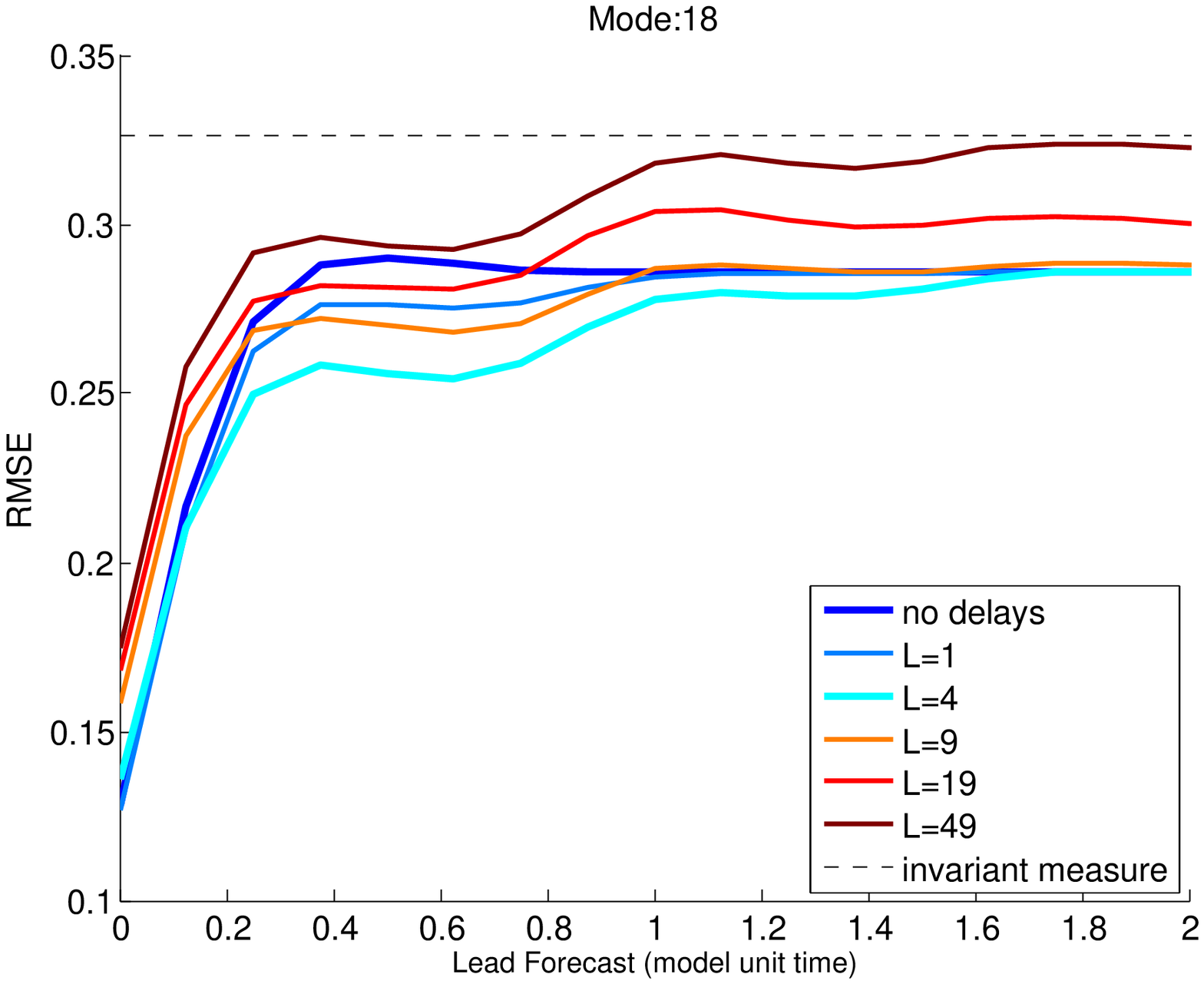}}
\end{center}
\caption{Forecasting skill (RMSE) of the diffusion model in predicting mode-8 (left) and mode-18 (right) of L-96 model with $F=6$ constructed using various choices of $L$.}
\label{L96_varyL}
\end{figure}

The nonparametric modeling method has two nuisance parameters, the number of lags $L$ in the time-delay embedding and the number of diffusion modes, $M$.  As discussed in Section \ref{sec22} a large number of lags can reduce the influence of independent identically distributed observation noise and also biases the geometry towards the stable component of the dynamics.  In the example in Section \ref{sec22} we saw that a large number of delays could significantly reduce the influence of observation noise on a one-dimensional dynamical system.  In Figure \ref{L96_varyL} we show that for higher-dimensional dynamical systems, there is a limit to the number of lags which can be used to reduce the influence of noise.  This limit is explained by the projection of the geometry onto the most stable component of the dynamics, which for large lags will practically eliminate less stable components which also contain information necessary for forecasting.  The number of lags that can be usefully included is limited by the size of the most stable Lyapunov exponent, since this controls the exponential bias of the delay-embedding towards the most stable component as shown in Section \ref{sec22}.  Notice that in Figure \ref{L96_varyL}, mode-18 appears to be more sensitive to the number of lags because the noise is much larger as a percentage of the variance.  In each case the standard deviation of the observation noise is $40^{-1/2} \approx 0.16$, and the standard deviation of the mode is shown by the RMSE of the corresponding invariant measure.  In each case the optimal number of lags is between $L=2$ and $L=10$, which is consistent with each mode being a generic observation of the same Lorenz-96 dynamical system, and hence each mode has the same most stable Lyapunov exponent. In the remaining numerical experiments in this section, we show results with $L=4$ lags in the time-delay embedding. 

\begin{figure}
\begin{center}
\mbox{
\includegraphics[width=.45\textwidth]{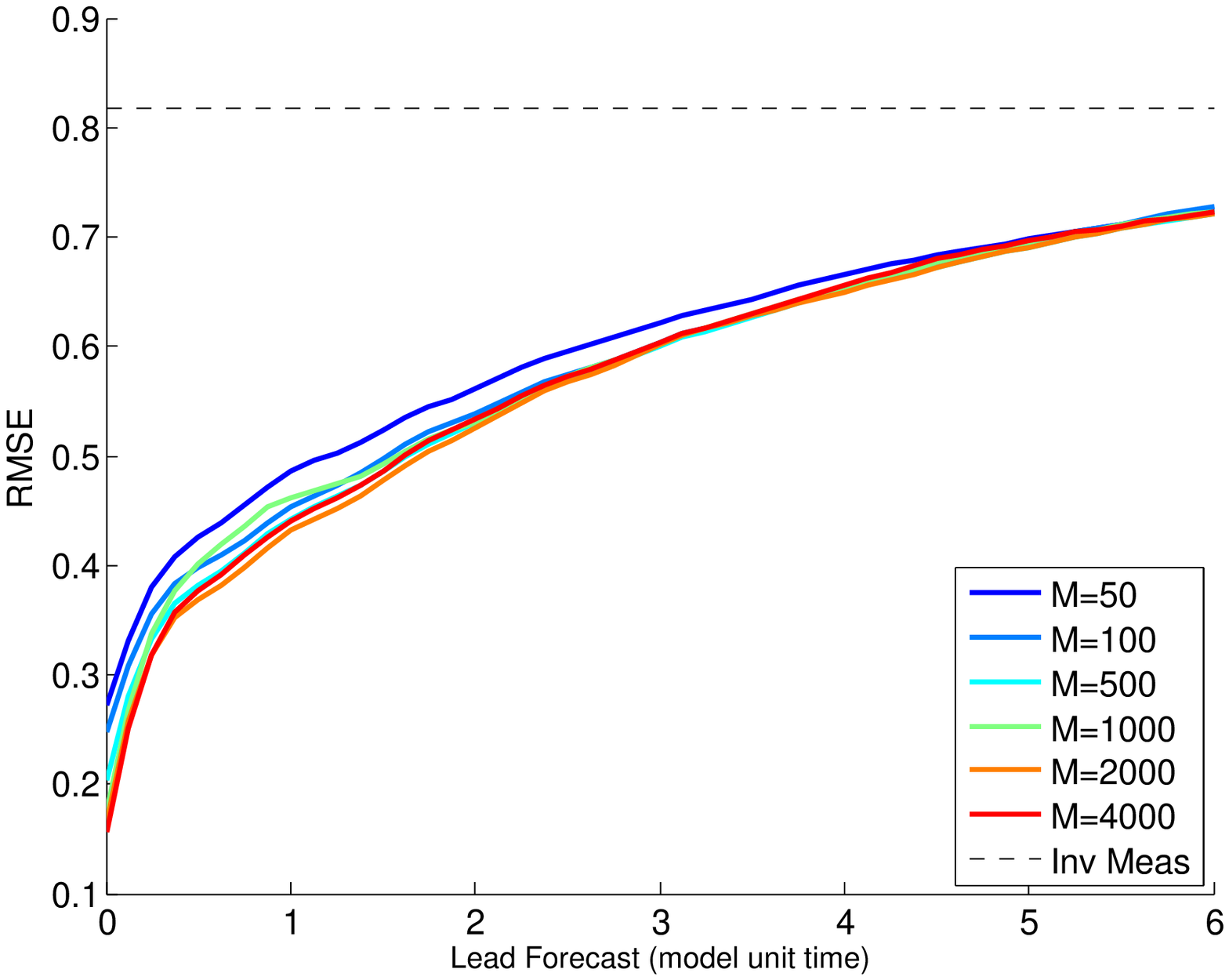}
\includegraphics[width=.45\textwidth]{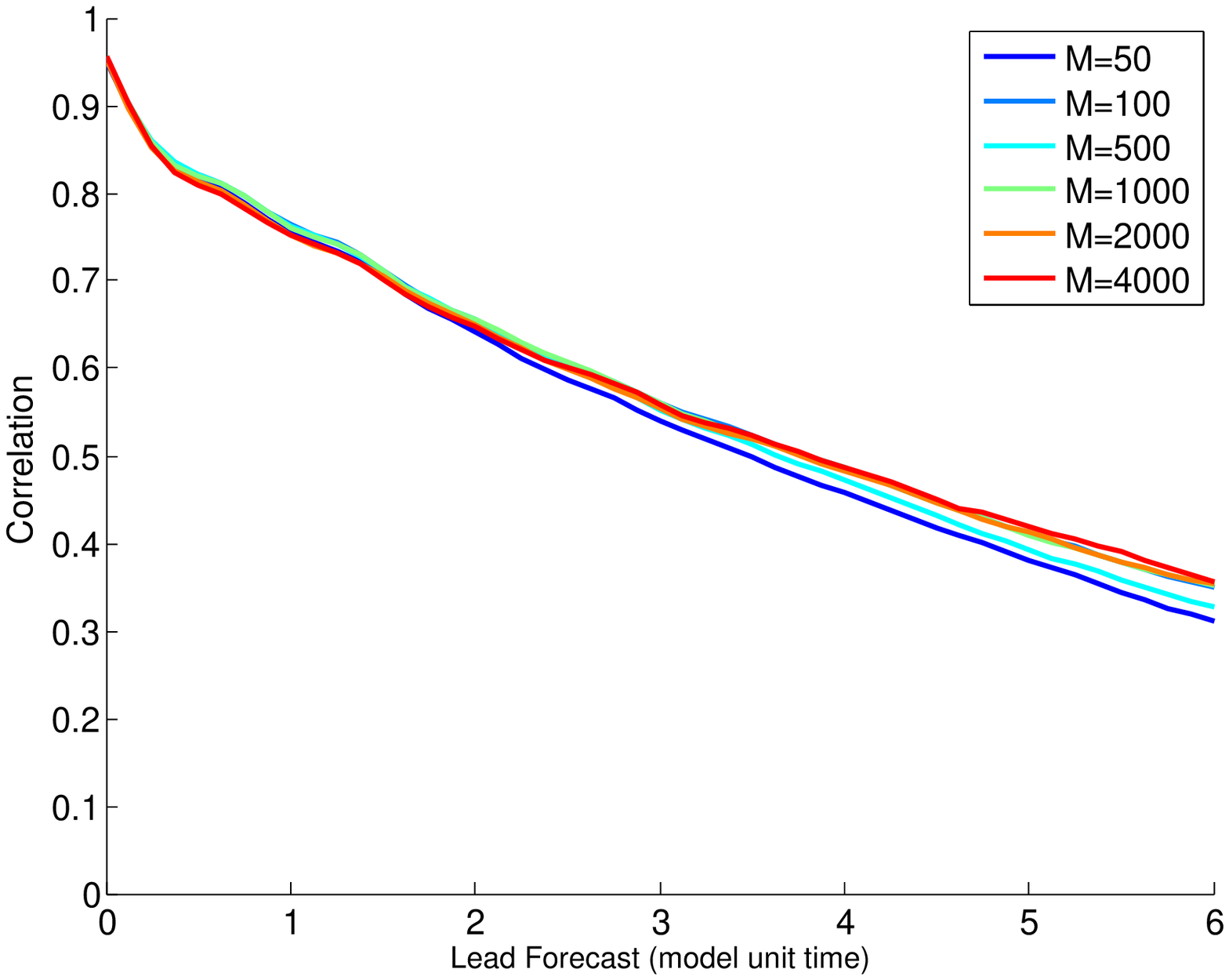}}
\end{center}
\caption{Forecasting skill of the diffusion model in predicting mode-8 of L-96 model with $F=6$ constructed using various choices of $M$.}
\label{L96_varyM}
\end{figure}

The other nuisance parameter is the choice of the number of diffusion modes, $M$. In Figure~\ref{L96_varyM} we show the forecasting skill for various choices of $M$ for an individual mode-8 for the case of weak turbulence $F=8$. Notice that while the initial errors increase as fewer modes are used, the forecast skill at later times are not very different. For deterministic problems, our experience suggests that larger $M$ help improve the prediction skill. For stochastic problems, however, the error estimate between $A$ and $\hat A$ derived in \cite{bgh:15} suggests that a large data set is required to obtain accurate estimation of the eigenfunctions associated with these high modes. At the moment, a systematic method for choosing $M$ is still under investigation and for the remainder of this paper, we simply set $M=2000$ for every mode. Figures \ref{L96_varyL} and \ref{L96_varyM} indicate that the diffusion forecast is not very sensitive to these nuisance parameters for turbulent modes, and these parameters can be tuned using cross-validation on the training data set. Also, there is clearly some improvement with appropriate choice of lag $L$ relative to no delay coordinate embedding. Even with such empirical choices of parameters ($L$ and $M$), we will see that the forecasting skill is still competitive to those of the perfect model when only the modeled modes are observed.

\subsection{Single-mode observations}

\begin{figure}
\begin{center}
\mbox{
\includegraphics[width=1\textwidth]{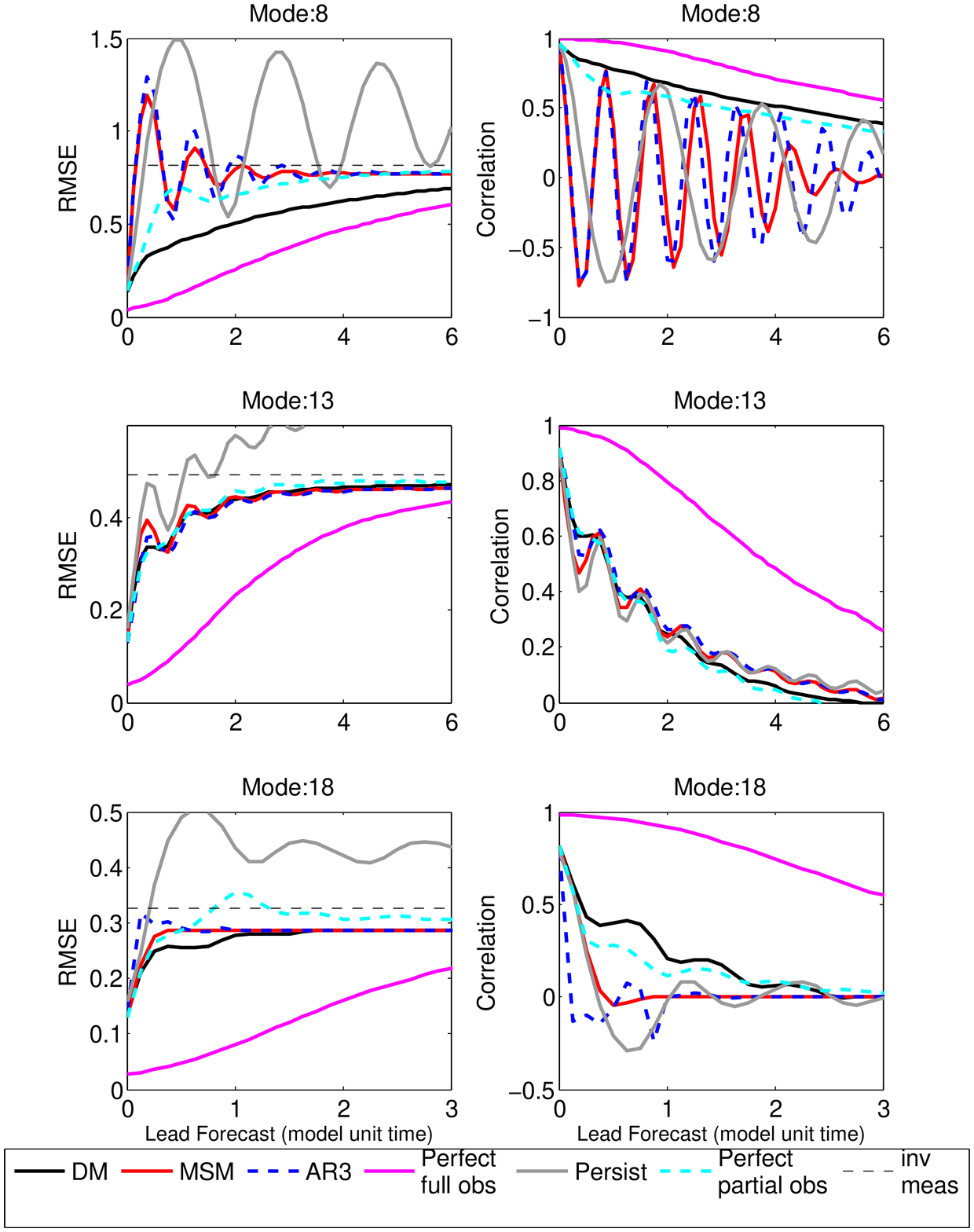}}
\end{center}
\caption{Forecasting skills of L96 models with $F=6$ for modes 8 (first row), 13 (second row), and 18 (third row) in terms of RMSE as defined in \eqref{rms} (first column) and pattern correlation in \eqref{pc} (second column).}
\label{L96_fig1}
\end{figure}

In Figure~\ref{L96_fig1}, we show the RMSE and pattern correlation for forecasting three different Fourier modes of the L96 model in a weakly chaotic regime with $F=6$: mode-8 carries the largest energy with a relatively short correlation time;  mode-13 carries low energy with a relatively long correlation time; and mode-18 carries low energy with a relatively short correlation time. Note that the ratios between the observation noise variance $\hat{R}=1/40$ and the energy of the modes are very different for these three modes; $4\%$ for mode-8, $11.7\%$ for mode-13, and $30\%$ for mode-18. Therefore, we expect that the noise does not affect the forecasting skill of modes-8 and 13 as much as that of mode-18. 

As expected, the perfect model with full observation provides the best forecast on all modes. The diffusion forecast performs very well on all modes, outperforming the perfect model when given equal information, meaning only observations of the corresponding individual mode (which is referred as ``perfect sparse obs" in Figure~\ref{L96_fig1}) on modes 8 and 18. For these modes, both autoregressive models (MSM, AR3) have very little predictive skill. On mode-13, the forecasting skill of all four models (AR-3, MSM, perfect sparse obs, and diffusion models) are not very different. The persistence model always produces severely biased forecasts in the long run, since fixing a particular observation will always be worse than predicting the statistical mean in a long term forecast. 

\begin{figure}
\begin{center}
\mbox{
\includegraphics[width=5in]{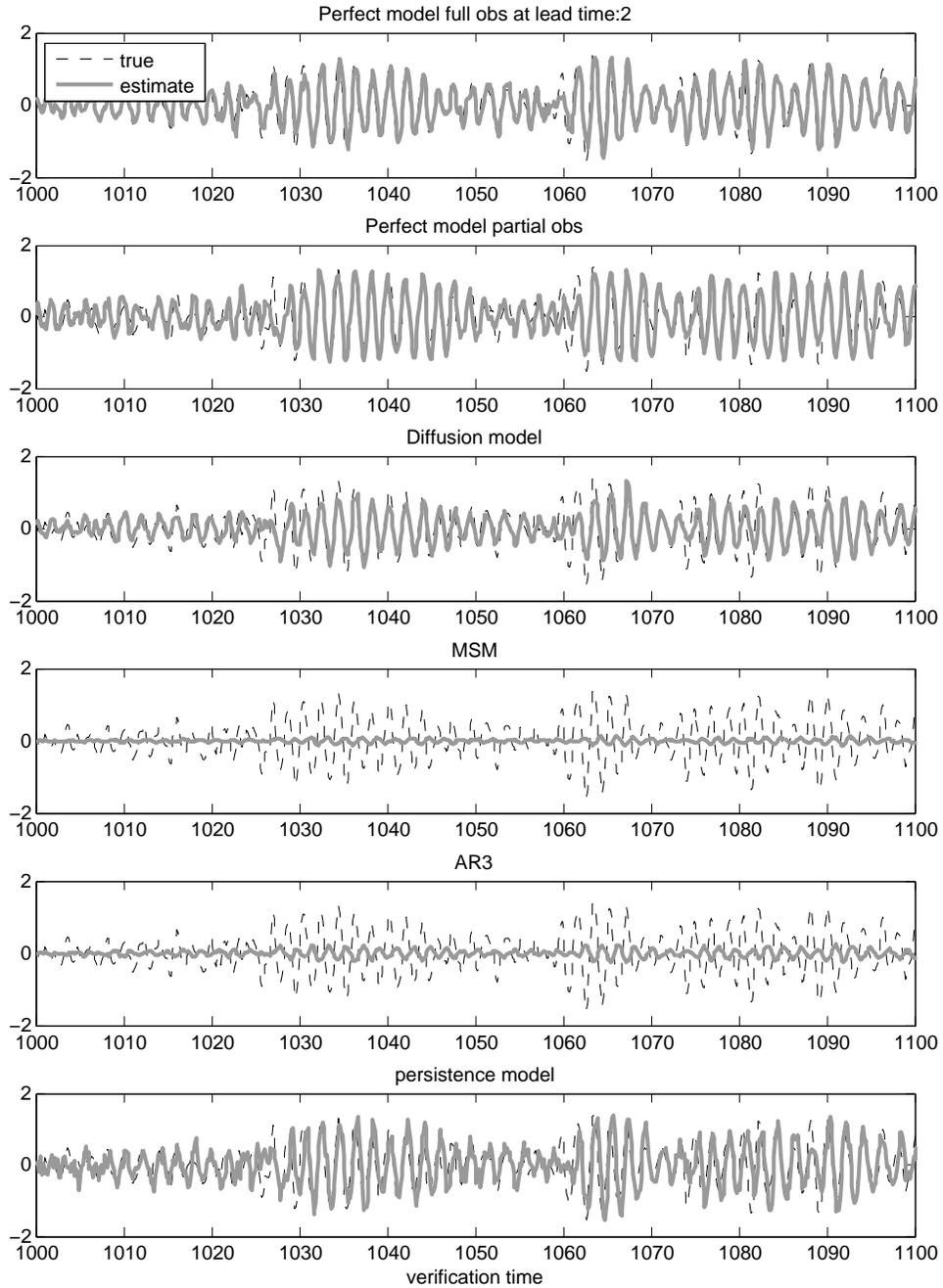}}
\end{center}
\caption{Mean forecast estimates for mode-8 of $F=6$ at lead times 2 model time unit at the verification period of $[1000,1100]$. The truth (black dashes) and estimates (gray solid).}
\label{L96_fig2}
\end{figure}

In Figures~\ref{L96_fig2}, we compare the forecast time series of each model to the truth for forecast lead time of 2 model time units. This comparison is made for mode-8 with $F=6$ for verification time steps between $1000$ and $1100$. Notice that the diffusion model forecast underestimates many of the peaks slightly whereas the perfect model initialized with the corresponding mode (perfect partial obs) overestimates many of the peaks. The persistence model forecasts are clearly out-of-phase. On the other hand, both the MSM and AR3 models significantly underestimate the amplitude of the signals and in the long term they simply forecast zero, as expected since these forecasts represent the mean of linear unbiased autoregressive models. 

\begin{figure}
\begin{center}
\mbox{
\includegraphics[width=1\textwidth]{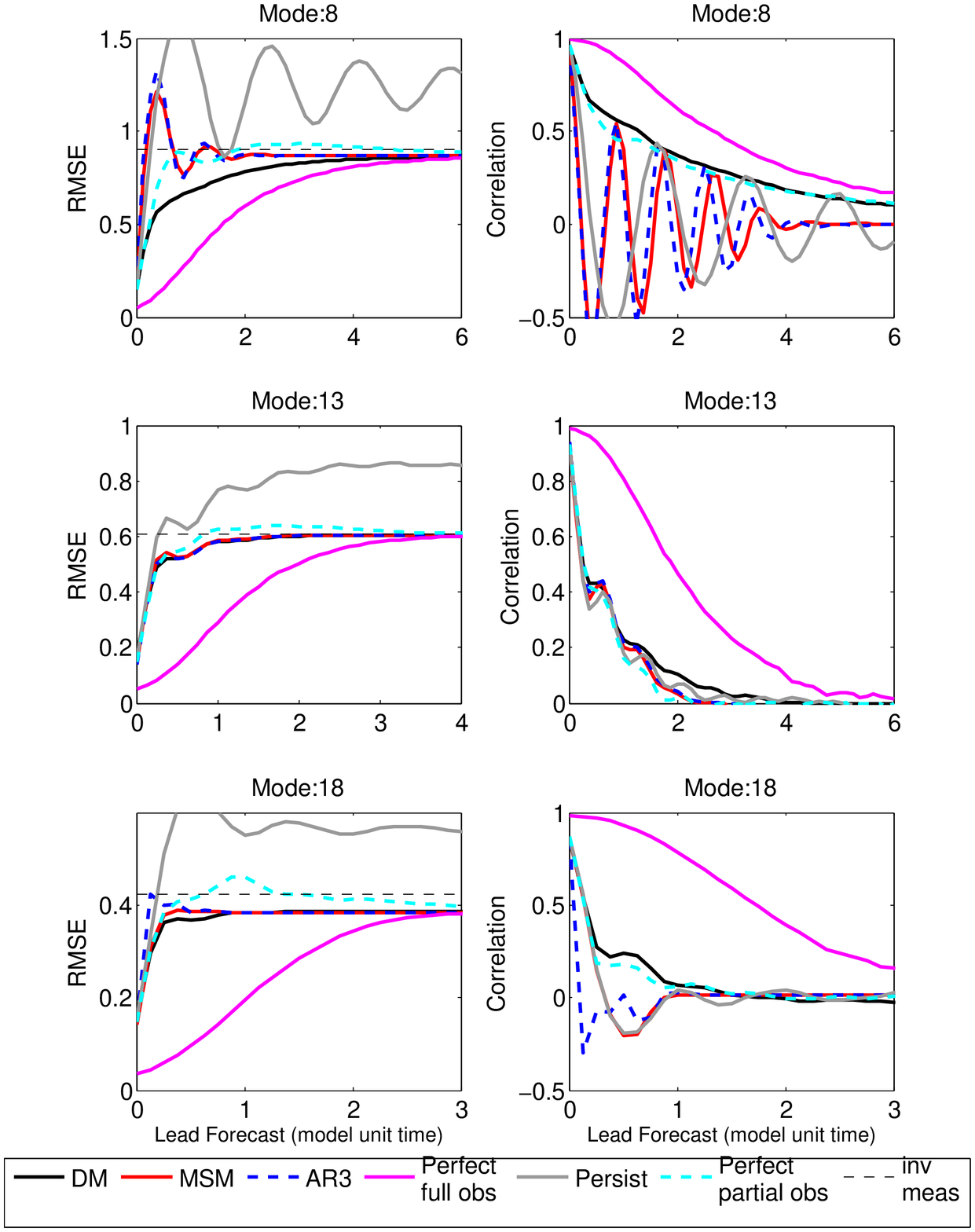}}
\end{center}
\caption{Forecasting skills of L96 models with $F=8$ for modes 8 (first row), 13 (second row), and 18 (third row) in terms of RMSE as defined in \eqref{rms} (first column) and pattern correlation in \eqref{pc} (second column).}
\label{L96_fig3}
\end{figure}

\begin{figure}
\begin{center}
\mbox{
\includegraphics[width=1\textwidth]{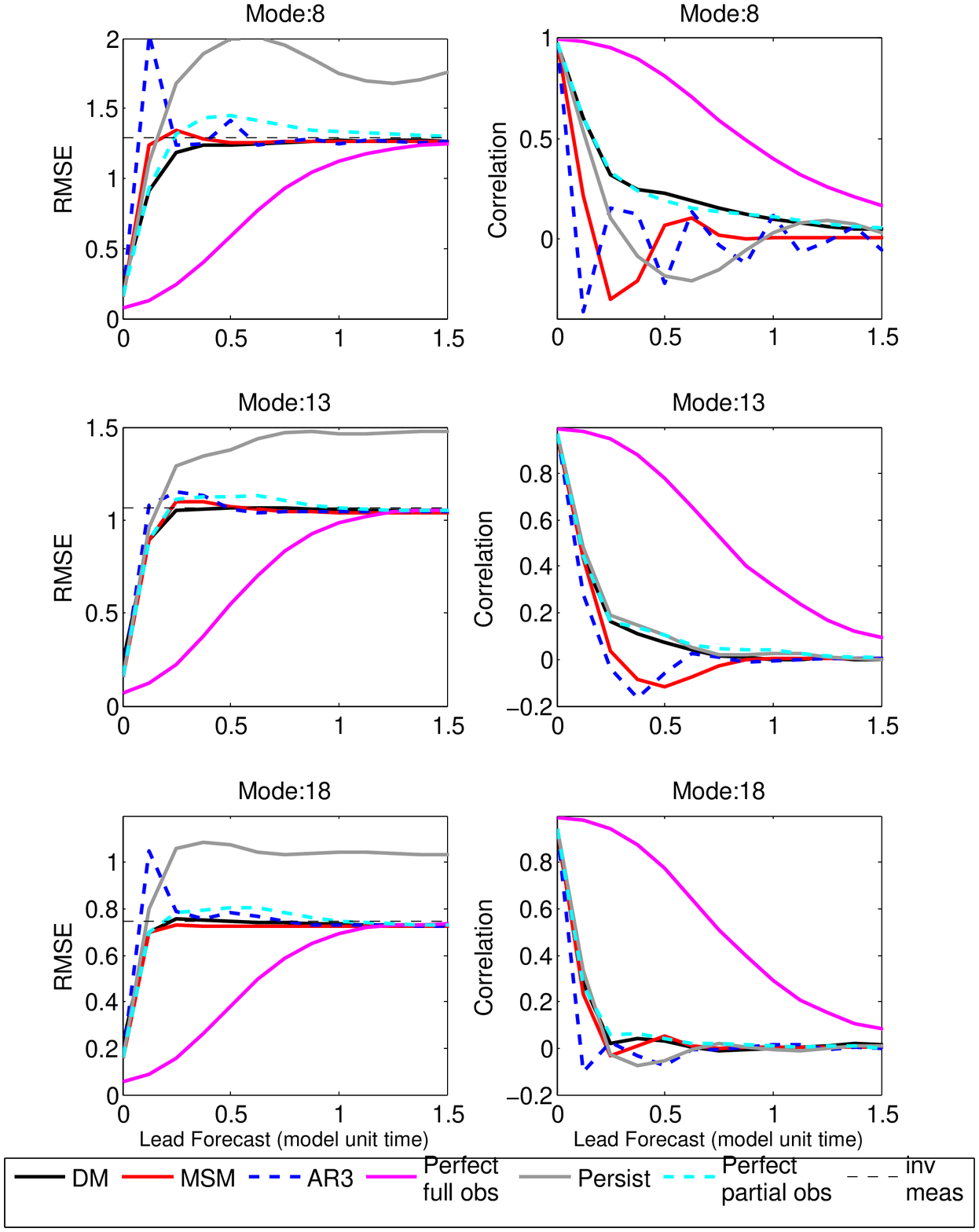}}
\end{center}
\caption{Forecasting skills of L96 models with $F=16$ for modes 8 (first row), 13 (second row), and 18 (third row) in terms of RMSE as defined in \eqref{rms} (first column) and pattern correlation in \eqref{pc} (second column).}
\label{L96_fig4}
\end{figure}

We also report the forecast skills of the more chaotic regime with $F=8$ in Figure~\ref{L96_fig3} and the fully turbulent regime with $F=16$ in Figure~\ref{L96_fig4}. Based on the RMSE and pattern correlation measures, these results suggest that the diffusion model still produces the most skillful forecasts compared to the MSM, AR3, the perfect model partial obs and persistence models on mode-8. On higher wave numbers, the forecasting skill diminishes as $F$ increases and the advantage of the nonparametric model over the parametric models become negligible.

The numerical experiments in this section suggest that the diffusion model is competitive with the perfect model when the available data is the corresponding noisy observations of the single mode that is being modeled. We also find that the diffusion model is most skillful when modeling the highly energetic modes. We should note that as $F$ increases, although mode-8 is still the most energetic mode, the energy is less dominant relative to the other modes in these turbulent regimes \cite{mag:05}. We suspect that this is one of the causes of degradation in the forecasting skill when $F$ is larger, in addition to the more quickly decaying correlation functions in the fully turbulent regimes.

\subsection{Multi-modes observations}
Now, we consider forecasting the three most energetic modes, ${\cal K}=\{7, 8, 9\}$. There are two strategies that can be adopted in the diffusion modeling: 1) Learning the 3-modes diffusion model directly; 2) Concatenate the three independent one-mode diffusion models (we refer to this strategy as Concatenated DM or CDM). Obviously, strategy 1) requires more data for accurate estimation since the dimensionality of the underlying dynamics that are being modeled increases. In our numerical experiment, however, we do not increase the amount of data, instead we use the same 5000 data points that were used in building the individual one-mode diffusion models.

\begin{figure}
\begin{center}
\mbox{
\includegraphics[width=1\textwidth]{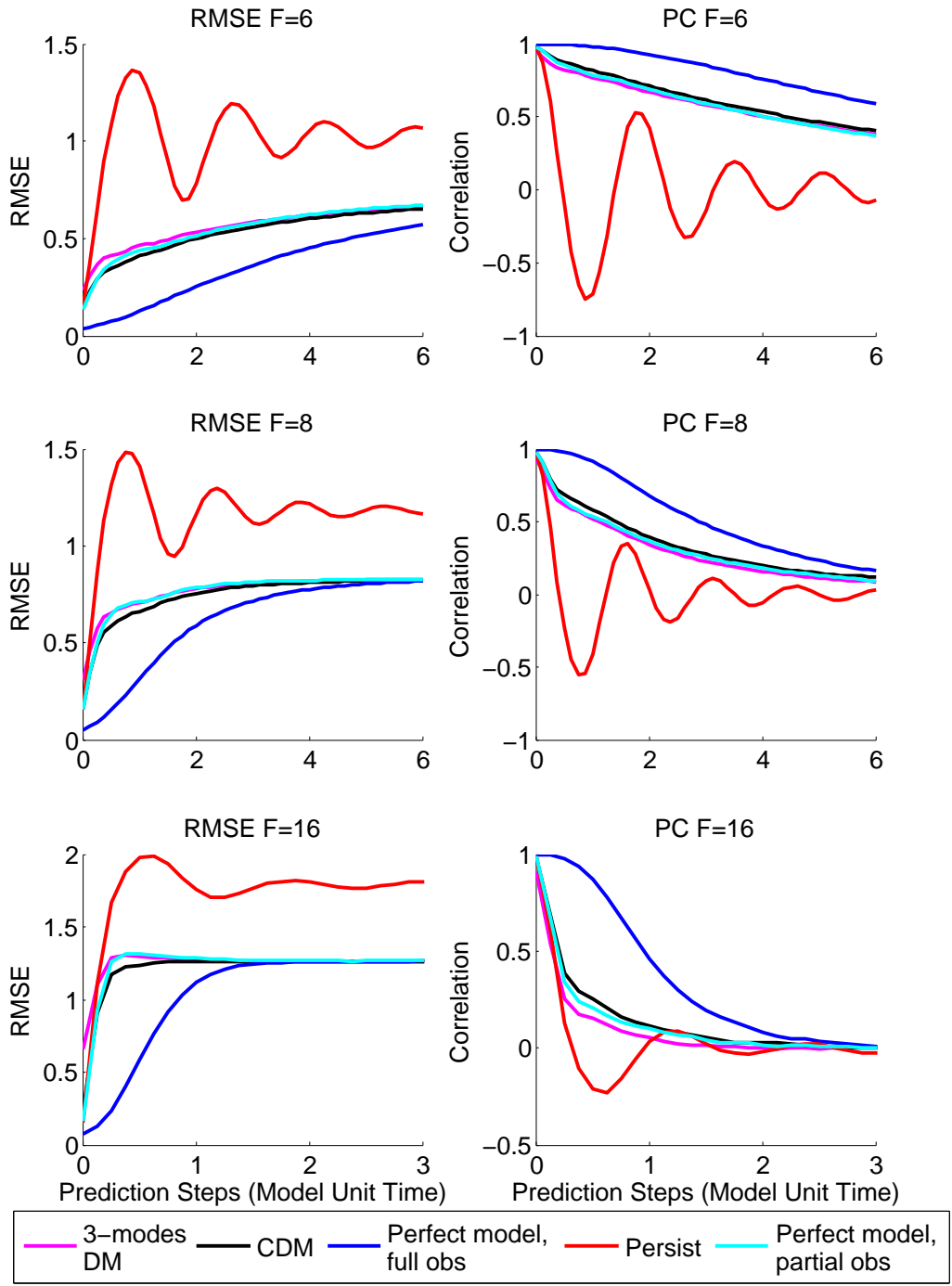}}
\end{center}
\caption{Forecasting skills of L96 models for the three most energetic mode $\{7, 8, 9\}$.}
\label{L96multi3modes}
\end{figure}

In Figure~\ref{L96multi3modes}, we see that while the perfect model given full data is still the best model, when only these three modes data are available, the forecasting skill of the perfect model is comparable to the concatenated three-modes diffusion model. The concatenated diffusion models (CDM) is slightly better than the three-modes DM since the underlying dynamics that are being modeled increases in CDM. In this case, these nonparametric approaches significantly supersede the persistence model.   

\begin{figure}
\begin{center}
\mbox{
\includegraphics[width=1\textwidth]{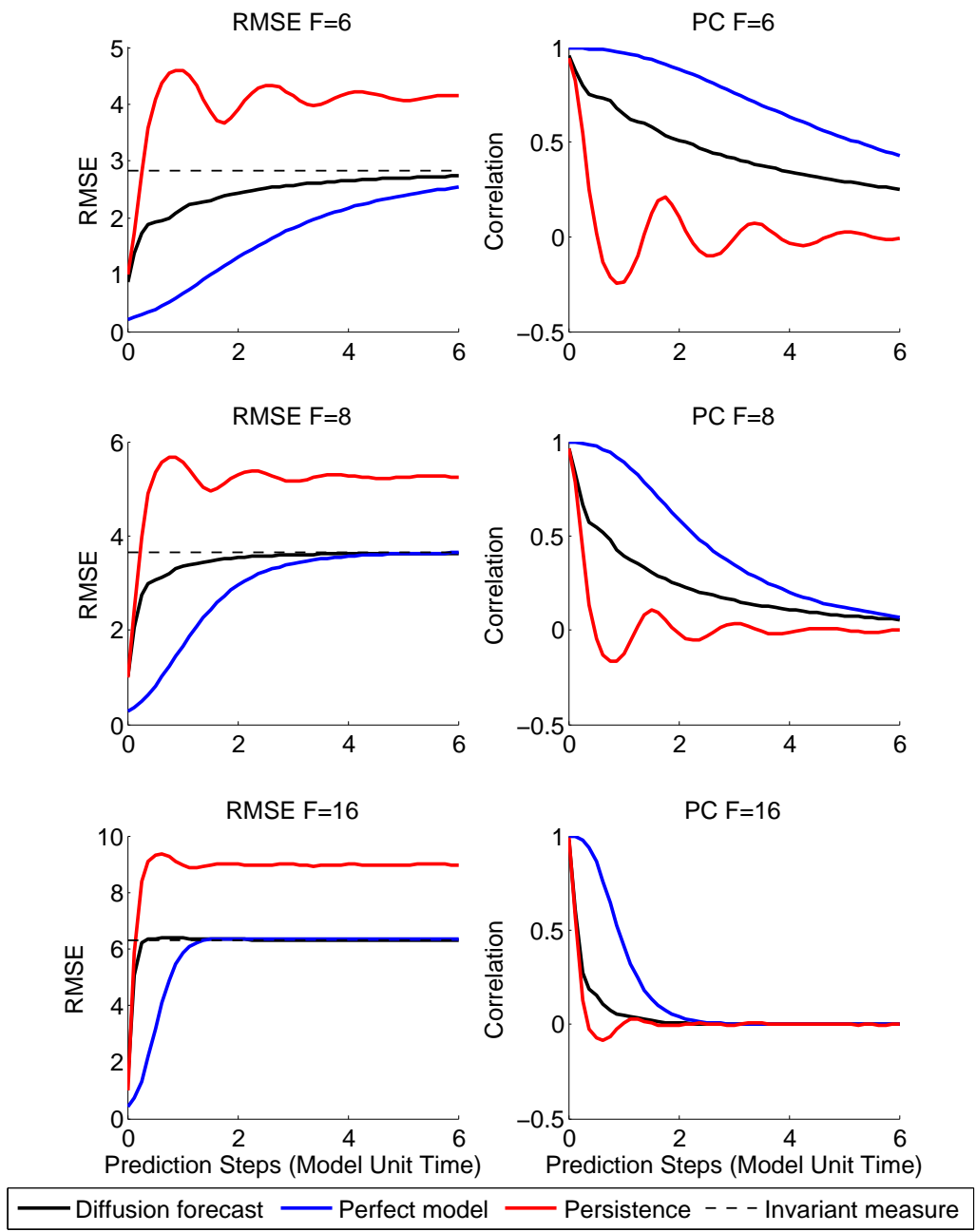}}
\end{center}
\caption{Forecasting skills of L96 models given full data.}
\label{L96spatial}
\end{figure}

Finally, we also report the overall performance of the diffusion forecast given data of all Fourier modes in Figure~\ref{L96spatial}. Since constructing the full 40-dimensional diffusion model is beyond the computational capacity, we just show results where we concatenate all 21 independently constructed one-mode diffusion models. In this verification, the quantifying RMSE \eqref{rms} and pattern correlation \eqref{pc} are measured in the spatial domain. While the diffusion forecast is not expected to be comparable to the perfect model given full observation data, one can see that it still produces some skillful forecast beyond the persistence model and the skill diminishes as $F$ increases. 

From these numerical experiments, one can see that if the only available data is an individual mode or few modes that are being predicted, the forecasting skill of the diffusion model is indeed competitive with that of the perfect model. The degrading performance of the perfect model is due to inaccurate estimation of initial conditions at unobserved modes. When the full data is available, this issue disappears in the perfect model experiment. Given all the observations, the nonparametric diffusion model does not have capability to represent the full nontrivial interactions of the dynamics that can only be achieved given the full model with accurate initial conditions. This degradation is totally expected since the nonparametric model is constructed with a short time series, and as we mentioned in the introduction, even if a large data set is available, the nonparametric model is subject to the curse-of-dimensionality. 

\section{Forecasting geophysical turbulent modes}

In this section, we consider forecasting turbulent modes of the quasigeostrophic model, a prototypical model for midlatitude atmosphere and oceanic dynamics \cite{salmon:98}. We consider a two-layer quasigeostrophic (QG) model which is externally forced by a mean shear with streamfunctions
\BEA
\Psi_1 = -Uy, \quad \Psi_2 = Uy,
\EEA
such that it exhibits baroclinic instabilities; the properties of the turbulent cascade has been extensively discussed in this setting (see e.g., \cite{salmon:98,vallis:06} and citations in \cite{smith:02}). 

The governing equations for the two-layer QG model with a flat bottom, rigid lid, and equal depth $H$ are given by,
\BEA
\left(\frac{\partial}{\partial t} +U\frac{\partial}{\partial x}\right)q_1+J(\psi_1,q_1)+\frac{\partial \psi_1}{\partial x}(\beta + k_d^2U)+\nu\nabla^8q_1=0,
\label{QG1}\\
\left(\frac{\partial}{\partial t} +U\frac{\partial}{\partial x}\right)q_2+J(\psi_2,q_2)+\frac{\partial \psi_2}{\partial x}(\beta - k_d^2U)+\kappa\nabla^2\psi_2+\nu\nabla^8q_2=0,
\label{QG2}
\EEA
where $\psi$ denotes the perturbed streamfunction, subscript 1 corresponds to the upper layer and subscript 2 corresponds to the bottom layer.  In \eqref{QG1}-\eqref{QG2}, $\beta$ is the meridional gradient of the Coriolis parameter; $\kappa$ is the Ekman bottom drag coefficient; $J(\psi,q)=\psi_xq_y-\psi_yq_x$ is a Jacobian function which acts as nonlinear advection; $U$ is the zonal mean shear, selected so that the QG equations exhibit baroclinic instability with a turbulent cascade; $\nu$ is the hyperviscosity coefficient, chosen so that $\nu\nabla^8q$ filters out the energy buildup on smaller scales when finite discretization is enforced.  The  perturbed QG potential vorticity $q$ is defined for each layer by 
\BEA
q_i=\nabla^2\psi_i+\frac{k_d^2}{2}(\psi_{3-i}-\psi_i).
\label{q}
\EEA The parameter $k_d = \sqrt{8}/L_d$ gives the wavenumber associated with radius of deformation, or Rossby radius, $L_d$ (the scale at which Earth's rotation becomes significant to the dynamics of the system).

In our numerical simulations, the true signal is generated by resolving (\ref{QG1})-(\ref{QG2}) with $128\times 64\times 2$ Fourier modes, which corresponds to the $128 \times 128 \times 2$ grid points. This model has two important nondimensional parameters:  $b=\beta(L/2\pi)^2/U_o,$ where $U_o=1$ is the horizontal non-dimensionalized velocity scale and $L$ is the horizontal domain size in both directions (we choose $L=2\pi$), and $F=(L/2\pi)^2/L_d^2,$ which is inversely proportional to the deformation radius \cite{hm:10b}. As in \cite{km:05,hm:10b}, we will consider two cases: one with a deformation radius $L_d$ such that $F=4$ which roughly mimics a turbulent jet in the midlatitude atmosphere and the other case will have a deformation radius $L_d$ such that $F=40$ which roughly mimics a turbulent jet in the ocean. For the ocean case, the QG model in \eqref{QG1}-\eqref{QG2} is numerically very stiff since the term that involves $k_d^2 = 8/L_d^2 = 8F$ is large.  

The large-scale components of this turbulent system are barotropic and for the two-layer model with equal depth, the barotropic streamfunction is defined as an average between the two layers, $\psi^b=\frac{1}{2}(\psi_1+\psi_2)$ \cite{salmon:98,km:05}. In the atmospheric regime with $F=4$, the barotropic flow $\psi^b$ is dominated by a strong zonal jet while in the ocean case with $F=40$ there are transitions between zonal flow and large-scale Rossby waves (e.g., see \cite{mfk:08,hm:10b,bh:13}). 

In our numerical examples, we build a diffusion model for the two-dimensional Fourier modes $\psi_{k,\ell}$ of the barotropic streamfunction, $\psi_b$. Following the experiments in \cite{hm:10b,bh:13}, our choice to only model this mode is mainly due to the fact that small-scale observations of a turbulent jet are typically not available, especially in the ocean. For diagnostic purposes, we compare the diffusion model to a simple stochastic model with combined white and colored additive and multiplicative noises that was designed as a test filter model for stochastic parameterization in the presence of model error \cite{ghm:10a,ghm:10b}. The governing equation of this model is given as follows,
\BEA
d\hat{\psi} &=& (-\gamma \hat{\psi} + \mathi\omega\hat{\psi} +b)\,dt + \sigma \,dW_{t},\nonumber\\
db &=& -(\lambda_b b-\bar{b})\,dt + \sigma_b\, dW_{b,t}, \label{spekf}\\
d\gamma &=& -(\lambda_\gamma \gamma - \bar{\gamma})\,dt +  \sigma_\gamma\, dW_{\gamma,t}\nonumber,
\EEA
where variable $\hat{\psi}$ models the Fourier mode of $\psi_{k,l}$, dropping the horizontal and vertical wave components $(k,\ell)$ to simplify the notation.  The equation governing $\hat{\psi}$ represents the interactions between this resolved mode and the remaining unresolved modes by several noise terms.  The first noise term, $\gamma$, is a real valued multiplicative noise. The second noise term is $b$, which is an additive complex valued noise governed by the Ornstein-Uhlenbeck mean reverting SDE.  The final noise term is $\sigma dW_t$, which is an additive white noise.  In \eqref{spekf}, the stochastic terms, $W_t, W_{b,t}$, are complex valued and $W_{\gamma,t}$ are real valued Wiener processes. We should note that while the simple system in \eqref{spekf} is nonlinear (due to the multiplicative term $-\gamma\hat{\psi}$), it has analytical statistical solutions \cite{ghm:10a,ghm:10b} which we will utilize in our numerical experiments here. 

In \cite{ghm:10a,ghm:10b,mh:12}, they implemented a Kalman update on the analytical mean and covariance solutions of \eqref{spekf} and called this filtering method SPEKF \cite{ghm:10a,ghm:10b,mh:12}, which stands for Stochastic Parameterized Extended Kalman Filter. In our implementation below, we will use SPEKF to generate initial conditions for our forecast at the verification times. As in any standard parametric modeling approach, one of the main issue is in choosing parameters $\{\omega,\sigma, \lambda_b,\bar{b},\sigma_b,\lambda_\gamma,\bar{\gamma},\sigma_\gamma\}$ in \eqref{spekf} to produce high forecasting skill. We use parameters that are tuned to optimize the filtered solutions as in \cite{hm:10b}. We found that this set of parameters produce better forecast compared to that obtained by an adaptive estimation method such as \cite{hmm:14} (results are not shown). Even though the parameters are empirically chosen, we shall see that the forecasting skill of the SPEKF model can be competitive with the diffusion forecast in some regimes. 

\begin{figure}
\begin{center}
\mbox{
\includegraphics[width=3in]{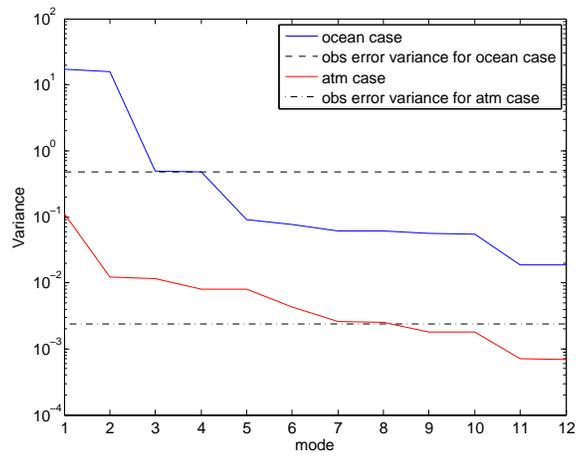}}
\end{center}
\caption{Observation error variance relative to the variances of the 12 Fourier modes corresponding to the 36 regularly spaced observed grid points.}
\label{energy}
\end{figure}

In our numerical experiments, we generate a time series of 9000 data points at discrete time step $\Delta t=1/4$ and added Gaussian noise in the physical space at 36 regularly spaced grid points in the spatial domain as in \cite{hm:10b} with noise variance $R=25\% Var(\psi^b)$. In Fourier space, the observation error variance is $\hat{R}=R/36$; in Figure~\ref{energy}, we show $\hat{R}$ relative to the variance of the 12 modes corresponding to the 36 regularly spaced observed grid points for both the atmospheric and ocean regimes. We use the first 5000 noisy data points to train the diffusion model and the remaining 4000 data points to compute the forecasting skill measures. For the diffusion forecast model, we empirically choose $L=9$ for the atmospheric regime and $L=4$ for the ocean regime. In all experiments, we use $M=2000$ diffusion modes. Note that one can compare the results with the perfect model forecasts, filtering the same 36 observations to generate initial conditions. In previous work \cite{hm:10b} it was shown that while the initial conditions for the atmospheric regime can be accurately specified with an appropriate EnKF scheme, in the ocean regime the initial conditions are not easily estimated due to numerical stiffness. Since finding accurate initial conditions for the perfect model is not the main point this paper, we neglect showing the perfect model experiment in this section. Instead, we just compare it with the truth which readily represents the best forecasting scenario.

\begin{figure}[t]
\begin{center}
\mbox{
\includegraphics[width=6in]{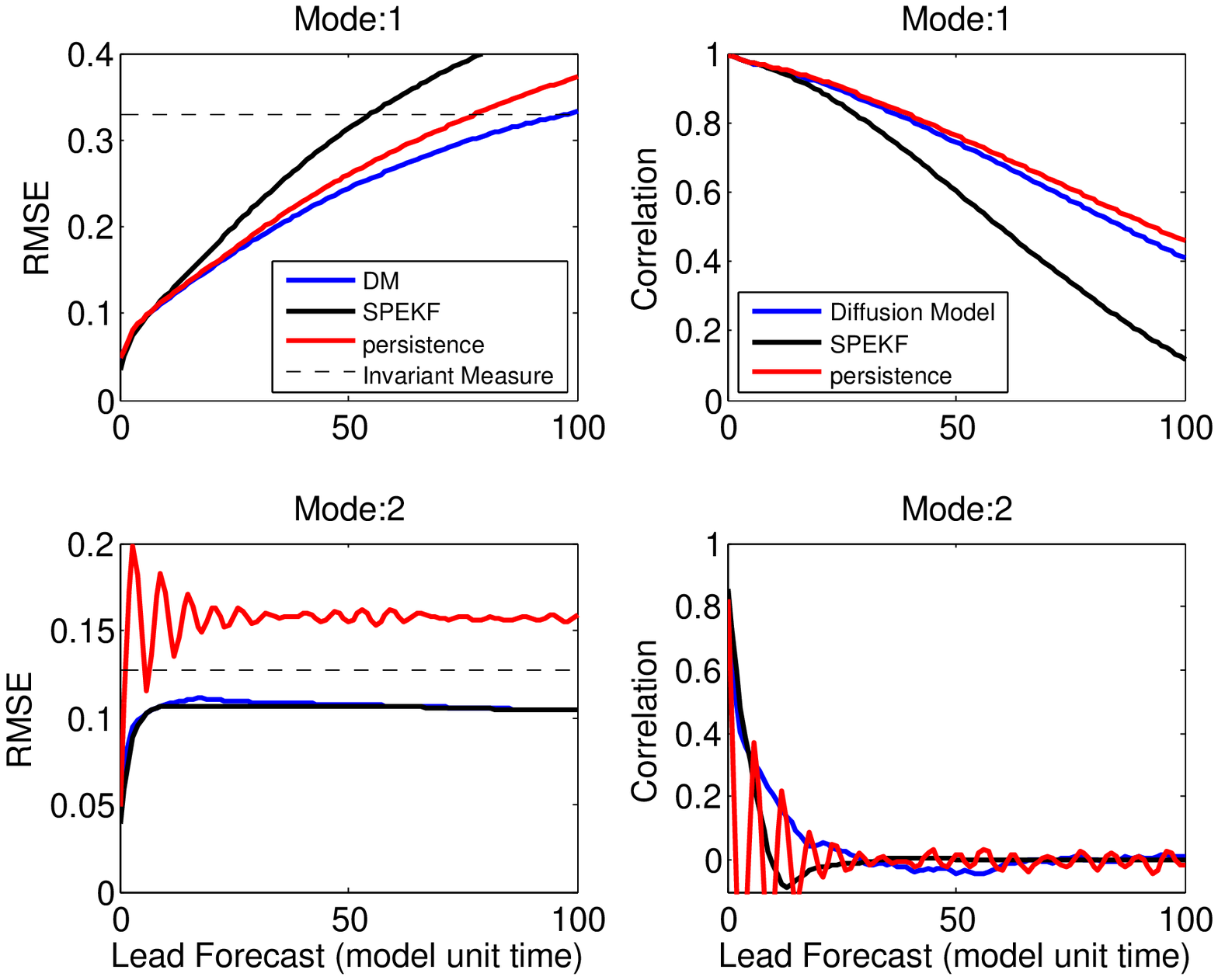}}
\end{center}
\caption{Forecasting skills of the QG models for the two most energetic modes for the atmospheric regime with $F=4$ in terms of RMSE as defined in \eqref{rms} (first column) and pattern correlation in \eqref{pc} (second column).}
\label{QG_fig1}
\end{figure}

\begin{figure}[t]
\begin{center}
\mbox{
\includegraphics[width=5in]{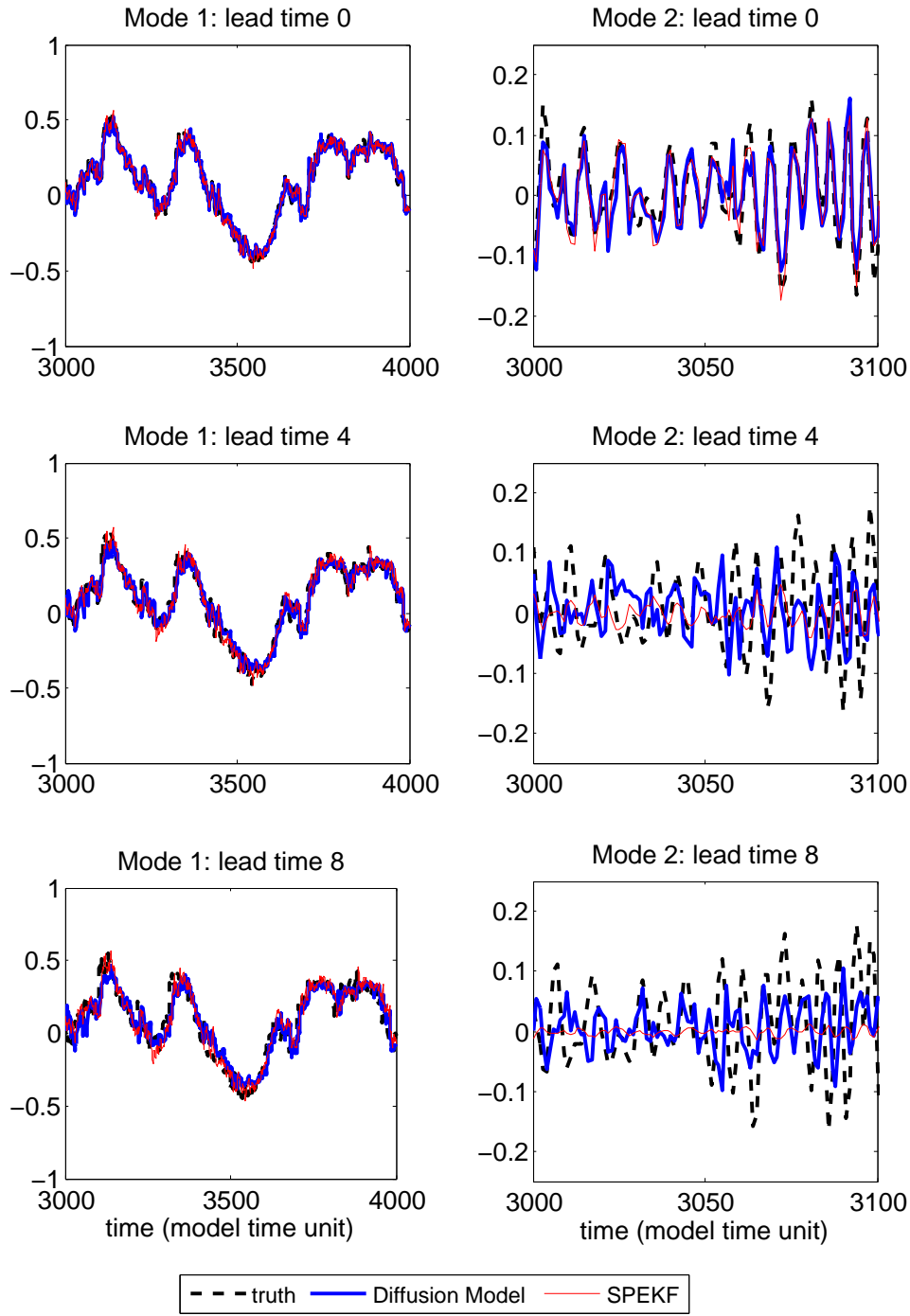}}
\end{center}
\caption{Forecasts of the QG models for the real component of the two most energetic modes for the atmospheric regime with $F=4$ 
at lead times $0, 4, 8$ on verification interval $[3000, 4000]$ for mode 1 and $[3000, 3100]$ for mode 2. }
\label{QG_fig2}
\end{figure}

In Figures~\ref{QG_fig1}, we show the RMSE and pattern correlation for the first two energetic modes in the atmospheric regime. Here, the first two energetic modes correspond to the two-dimensional horizontal wave numbers $(0,1)$ and $(1,1)$ with explained variances of about 65\% and 73\%, respectively, so the behavior is dominated by the zonal jet in the horizontal direction \cite{hm:10b}. In this regime, notice that the persistence model produces high forecasting skill with a very long correlation for the first mode. However, the persistence model is not useful in forecasting the second mode. In contrast, the parametric SPEKF model produces biased long-time forecast on the first mode but more accurate forecast on the second mode. In this regime, the diffusion forecasting skill is quite robust for both modes with best RMSE and PC scores. In Figure~\ref{QG_fig2}, we show the corresponding forecast time series of the real component of these two most energetic modes at lead times 0, 4, and 8 model time units. Notice that the Bayesian filter works reasonably well, indicated by the accurate recovery at lead time-0. The forecast skills of SPEKF and diffusion forecast for the first mode are relatively similar and very skillful at lead times 4 and 8 with PC high scores, above 0.9. For the second mode, on the other hand, one can see the forecasts of both models become less accurate at these lead times with very low PC, less than 0.5. Here, notice that SPEKF forecasts are nearly zero.

\begin{figure}[t]
\begin{center}
\mbox{
\includegraphics[width=6in]{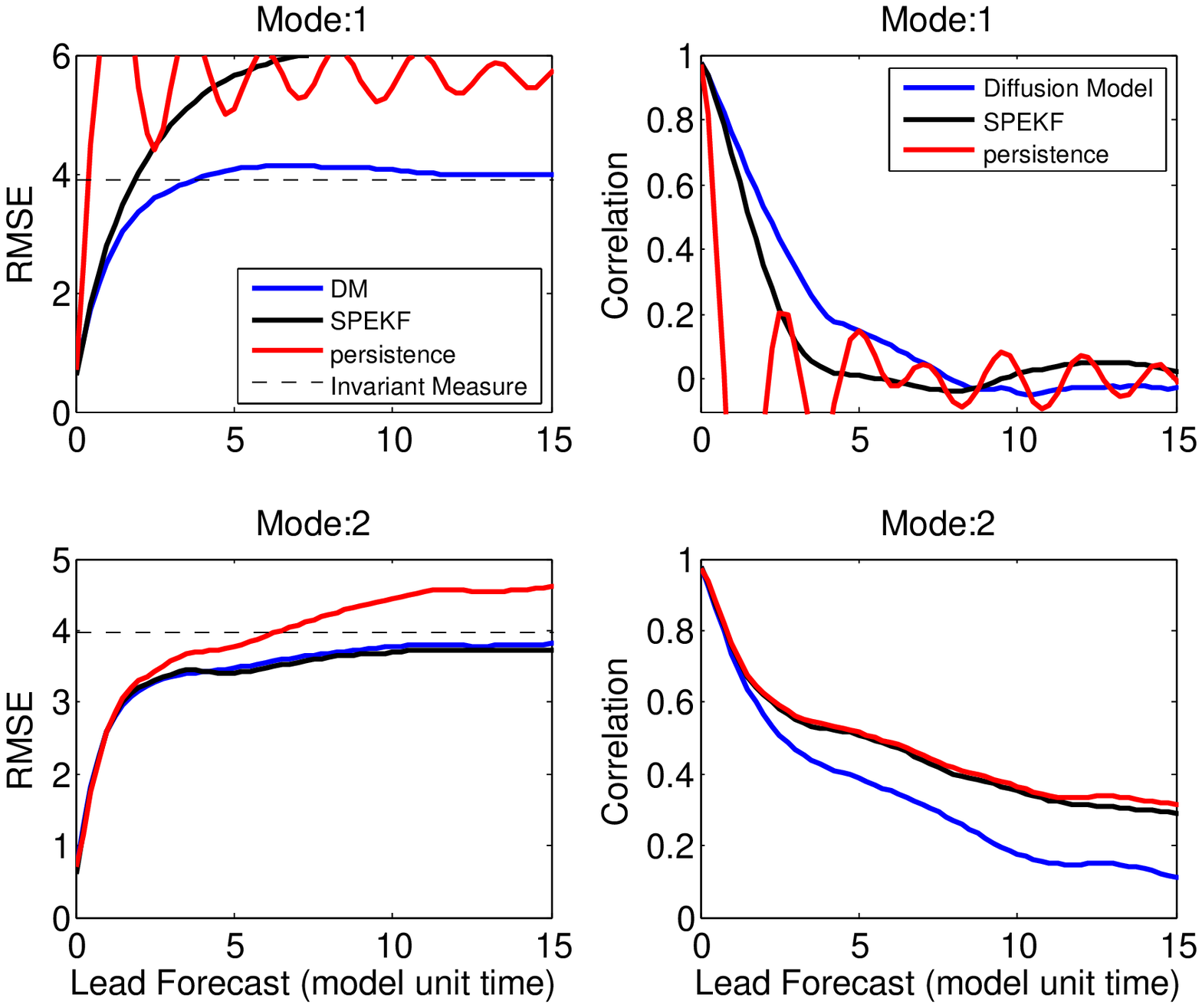}}
\end{center}
\caption{Forecasting skills of the QG models for the two most energetic modes for the ocean regime with $F=40$ in terms of RMSE as defined in \eqref{rms} (first column) and pattern correlation in \eqref{pc} (second column).}
\label{QG_fig3}
\end{figure}

\begin{figure}[t]
\begin{center}
\mbox{
\includegraphics[width=5in]{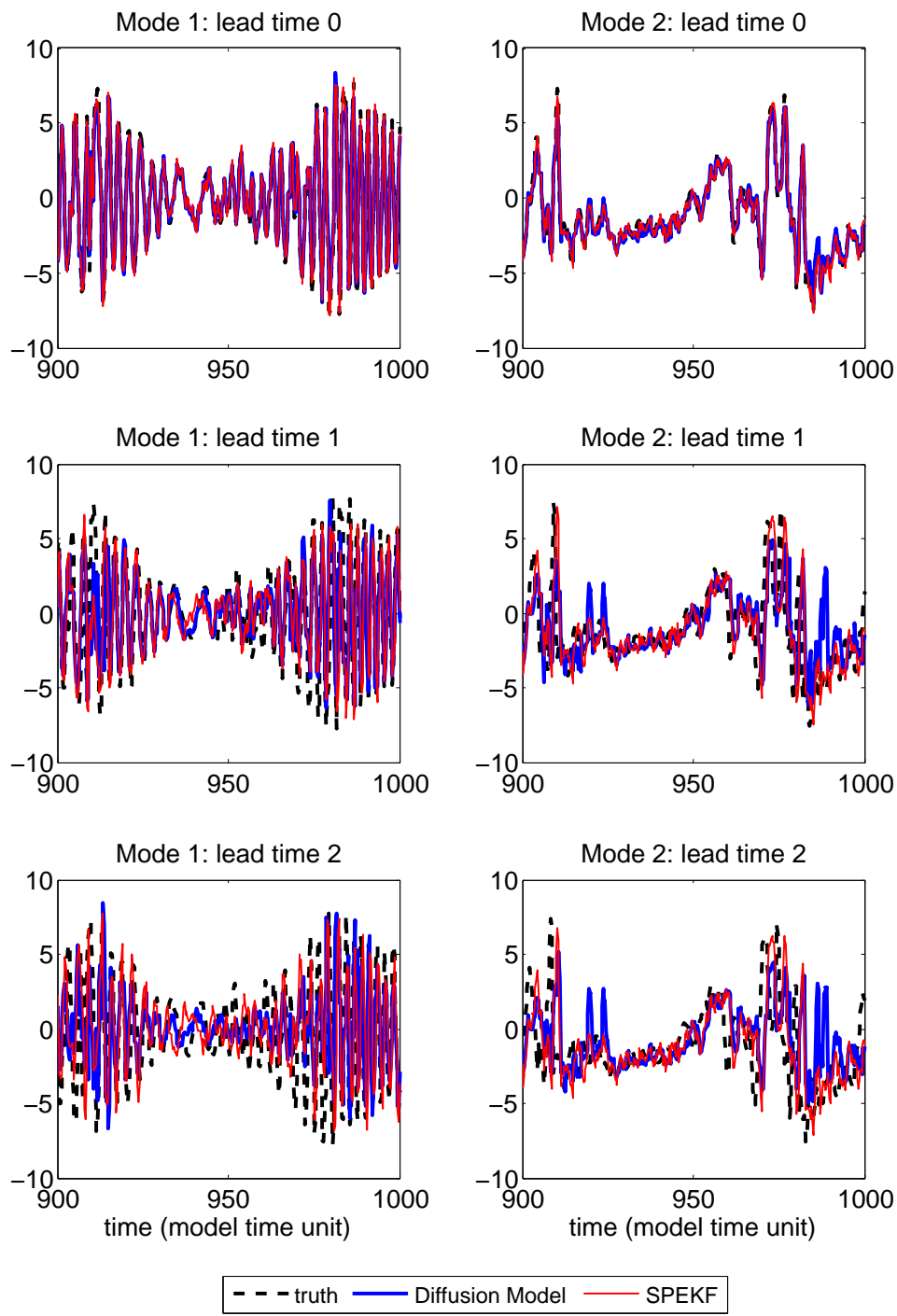}}
\end{center}
\caption{Forecasts of the QG models for the real component of the two most energetic modes for the ocean regime with $F=40$ 
at lead times $0, 1, 2$ on the verification interval $[900,1000]$.}
\label{QG_fig4}
\end{figure}

In Figures~\ref{QG_fig3}, we show the RMSE and pattern correlation for the first two energetic modes in the ocean regime. Here, the first two energetic modes correspond to the two-dimensional horizontal wave numbers $(1,0)$ and $(0,1)$ with explained variance of about 50\% and 97\%, respectively, so the behavior is dominated by two competing modes, the Rossby mode $(1,0)$ and the zonal jet mode $(0,1)$ \cite{hm:10b}. In this regime, the persistence model forecast has no skill at all on the first mode. On the second mode, however, the persistence forecast is quite reliable in the short term. The diffusion forecast is better than SPEKF on the first mode, but slightly worst than SPEKF on the second mode (indicated by worse PC score beyond lead time 2). In Figures~\ref{QG_fig4}, we show the corresponding forecast time series of the real component of these two most energetic modes for lead times 0, 1, and 2. At these lead times, where the PC score is at least 0.6, we expect the forecasting skill to be comparable on the first mode while SPEKF should be slightly better than the diffusion forecast on the second mode. 

\begin{figure}[t]
\begin{center}
\mbox{
\includegraphics[width=5in]{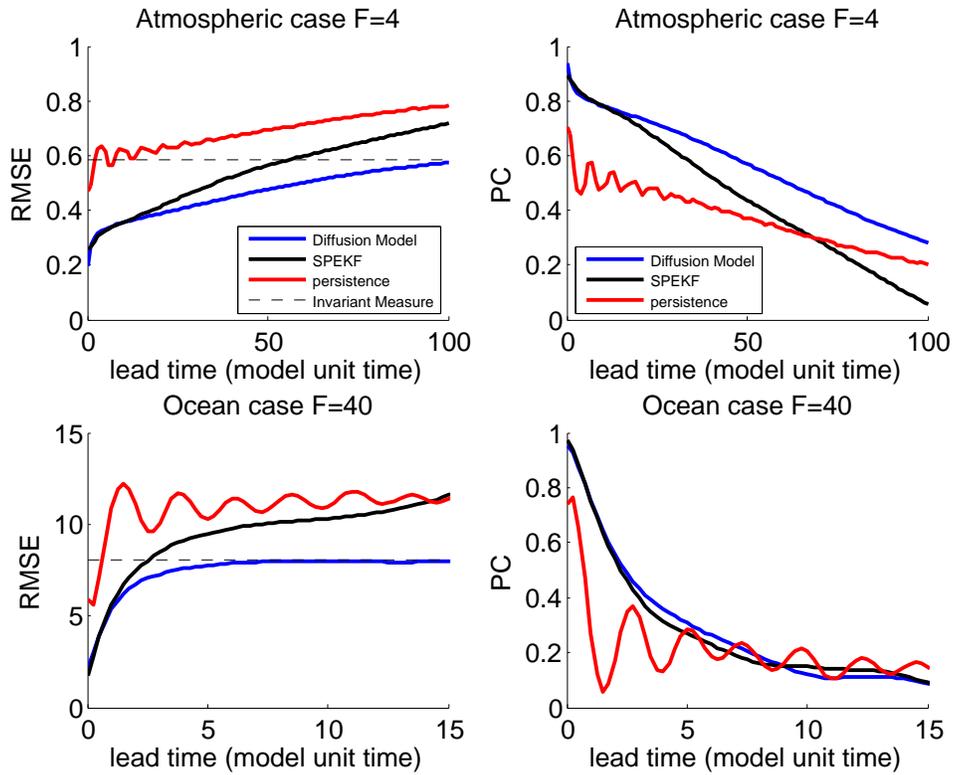}}
\end{center}
\caption{RMSE and PC scores for the atmospheric (top) and ocean (bottom) regimes, computed over the $6\times 6$ observed grid points and 4000 verification times.}
\label{QG_fig5}
\end{figure}

In Figure~\ref{QG_fig5}, we report the RMSE and pattern correlation computed over the $6\times 6$ observed grid points; obtained by inverse Fourier transforming 12 independent models for the corresponding modes. In terms of filtering, notice that the estimates from both the SPEKF and diffusion models are much more accurate compare to the estimates from the persistence model whose error is nothing but the observation noise standard deviation. In both regimes, the overall performances in terms of RMSE and PC scores from the best to the worst are the Diffusion model, SPEKF, and the persistence model. Notice that the forecasting skill of the diffusion model and SPEKF are competitive in the very short term forecast. In the long term, however, while the RMSE of the diffusion forecast always converges to the climatological standard deviation, SPEKF RMSE exceeds this climatological standard deviation, indicating a biased forecast. From the simulations in this section on a nontrivial prototypical example of midlatitude geophysical turbulent flows, we see that the proposed non-parametric diffusion models produce robust, skillful, and unbiased long term forecasts.   

\section{Summary}

In this paper, we applied the nonparametric diffusion modeling approach proposed in \cite{bgh:15} to forecast Fourier modes of turbulent dynamical systems given a finite set of noisy observations of these modes. Motivated by the results in \cite{DMDC,giannakisMajda,GiannakisPNAS}, we built the nonparametric model in the delay embedding coordinates in order to account for the interaction between the resolved and unresolved modes as well as to smooth out the noisy observations.  We empirically verify that larger $L$ helps to reduce the influence of the observation noise, however our theoretical derivation suggests that taking $L$ too large will project away less stable components of the dynamics. The choice of $L$ should not be so large as to project away the unresolved modes, but large enough to smooth out the noise in the training data set, and currently this selection requires empirical tuning. To initiate each forecast, we introduced a simple Bayesian filtering method which accounts for the information of the noisy observations up to the current time. We compared the forecasting skills of the proposed nonparametric model with various standard modeling approaches when the underlying dynamics are unknown, 
including the persistence model, linear autoregressive models of order-1 (MSM) and order-3, the SPEKF model which involves combined additive and multiplicative noise terms. We also compared the diffusion forecast to the perfect model forecast in the L96 example above. 
In this example, we found that when the only available observations are the low-dimensional set of the modes that are being modeled, then the diffusion forecasting method is competitive with the perfect model. 
When full data (all modes) becomes available, the curse-of-dimensionality limits the ability of diffusion model to represent nontrivial interactions of the underlying dynamics that can only be achieved given the perfect model and accurate initial conditions. Overcoming this limitation of the diffusion forecast is a significant challenge which we hope to address in future work.
From the QG experiments, we found that while the short-term forecast of the parametric model, SPEKF, is competitive with that of the diffusion model, the latter produces more robust and unbiased long term forecasts.  

One important thing to take away from this paper is that although the proposed nonparametric model seems to be competitive with the standard parametric models simulated here and even with the perfect model with partially observed data, we do not claim that this equation-free approach is the model to use for everything. If one knows the physics of the underlying dynamics (or the appropriate reduced parametric models), then one should use the physics-based model rather than the equation-free diffusion models. Indeed, it was shown rigorously in a simple setup that optimal filtering and accurate statistical prediction can be achieved with an appropriate stochastic parametric modeling of the unresolved scales \cite{PRSA}. However, there are at least two valuable aspects of the nonparametric model we develop here. First, one can use this model to diagnose whether his/her modeling approach is appropriate; we expect that appropriate physics-based models should beat this black box approach. Second, when an appropriate, physically motivated model is not available, then this approach will often outperform adhoc parametric models as demonstrated in this paper. Moreover, in practice it is difficult to guess the appropriate choice of parametric models even if some physics-based knowledge has been imposed in deriving a reduced model (e.g., in turbulent closure problems, one typically uses parameters to represent some small-scale processes or some physical processes which are not well understood). In this situation, we propose to extract the time series of these processes (if possible) and to subsequently build nonparametric models for such processes. This idea is what we refer to as the \emph{semi-parametric modeling} approach and is introduced in the companion paper \cite{bh:15semi}.

\section*{Acknowledgment}
The research of JH was partially supported by the Office of Naval Research Grants N00014-13-1-0797, MURI N00014-12-1-0912, and the National Science Foundation grant DMS-1317919. TB was partially supported through ONR MURI grant N00014-12-1-0912.

\section*{References}

\section*{Appendix A. Variable Bandwidth Diffusion Map Algorithm}

For data $\{x_i\}_{i=1}^N$ lying on a smooth manifold $\mathcal{M}$ (with $L=0$ lags), or for data given by a generic observation of a dynamical system on a smooth manifold (with $L$ sufficiently large), the algorithm given below will provably estimate the eigenfunctions $\varphi_j$ of the generator $\hat{\mathcal{L}}$ of \eqref{gradflow}.  The estimate $\hat {\bf L}$ of the operator
\begin{align}
\hat{\mathcal{L}} = -\nabla U\cdot \nabla + \Delta,\nonumber
\end{align}
where $U = -\log(\peq)$, is up to a pointwise error of order $\mathcal{O}\left(\epsilon, \frac{q(x_i)^{(1-d\beta)/2}}{\sqrt{N}\epsilon^{2+d/4}} ,\frac{||\nabla f(x_i)||q(x_i)^{-c_2}}{\sqrt{N}\epsilon^{1/2+d/4}} \right)$, where $q(x)$ denotes the sample distribution, which by ergodicity is exactly the invariant measure $\peq$ of the underlying dynamics.  We note that $c_2 = d(1/4-d/2) < 0$ for the choice $\beta = -1/2$ and $\alpha = -d/4$ so that the final error term is bounded even when $q(x_i)$ is arbitrarily close to zero.  For a derivation of these facts see \cite{bh:15vb} and for a brief overview see the supplemental material of \cite{bgh:15}. Below, we only give a step-by-step cookbook for constructing the eigenfunctions $\{\varphi_j\}$ of $\hat{\cal L}$.

\begin{enumerate}
	\item Choose delay weight $\kappa>0$ and number of lags $L$.  In our numerical experiments, we use $\kappa=0$.\vspace{2pt}
	\item Let $x_i = x(t_i) \in \mathbb{R}^n$ be a time series with $i=1,...,N+L$ data points. In our case, the data set are noisy Fourier modes $\hat{v}_{i,k}$.
  	\item For $i=1+L,...,N+L$ form the state vector, \[ y_i = [x_i,e^{-\kappa}x_{i-1},...,e^{-L\kappa}x_{i-L}]^T \in \mathbb{R}^{n(L+1)}.\] 
	\item For each $i$ find the $k$-nearest neighbors of $y_i$ in ${\mathbb R}^{n(L+1)}$, let their indices be $I(i,j)$ for $j=1,...,k$ ordered by increasing distance.  Here we used $k=512$. \vspace{2pt}
	\item Form a sparse $(N-L) \times (N-L)$ matrix with $(N-L)k$ nonzero entries given by \[ d(i,I(i,j)) = ||y_i - y_{I(i,j)}||. \]
	\item Define the ad hoc bandwidth function $\hat\rho_i = \sqrt{\sum_{j=1}^8 d(i,I(i,j))^2}$.\vspace{2pt}
	\item Automatically tune the bandwidth for the kernel density estimate.\vspace{2pt}
	\begin{enumerate}
		\item Let $\epsilon_l = 2^l$ for $l = -30,-29.9,...,9.9,10$.\vspace{2pt}
		\item Compute $T_l = \sum_{i,j=1}^N \exp\left(\frac{-d(i,j)}{2\epsilon_l \hat\rho_i \hat\rho_j}\right)$.\vspace{2pt}
		\item Estimate the local power law $T_l = \epsilon_l^a$ at each $l$ by $a_l = \frac{\log T_l - \log T_{l-1}}{\log \epsilon_l - \log \epsilon_{l-1}}$.\vspace{2pt}
		\item Choose $\epsilon = \textup{argmax}_{\epsilon_l}\{a_l\}$.\vspace{2pt}
		\item Estimate the intrinsic dimension $d = 2\max_{\epsilon_l} \{a_l\}$.\vspace{2pt}
	\end{enumerate}
	\item Form the density estimate $q_i = q(x_i) = (2\pi\epsilon \hat\rho_i^2 )^{-d/2}N^{-1}\sum_{j=1}^N \exp\left(\frac{-d(i,j)}{2\epsilon \hat\rho_i \hat\rho_j}\right)$.	\vspace{2pt}
	\item Define the bandwidth function $\rho_i = q_i^\beta$.  We use $\beta = -1/2$ and $\alpha = -d/4$.\vspace{2pt}
	\item Repeat step 7 to choose the bandwidth $\epsilon$ with $T_l = \sum_{i,j=1}^N \exp\left(\frac{-d(i,j)}{4\epsilon_l \rho_i \rho_j}\right)$.\vspace{2pt}
	\item Form the sparse kernel matrix $K(i,I(i,j)) = \exp\left(\frac{-d(i,I(i,j))}{4\epsilon \rho_i \rho_{I(i,j)}}\right)$.\vspace{2pt}
	\item Form the symmetric matrix $K^S = (K + K^T)/2$.\vspace{2pt}
	\item Form the diagonal normalization matrix $D_{ii}= \sum_{j=1}^N K^S_{ij}/q_i^{d\beta}$.\vspace{2pt}
	\item Normalize to form matrix $K^S_{\alpha}= D^{-\alpha} K^S D^{-\alpha}$.\vspace{2pt}
	\item Form the diagonal normalization matrix $(D_{\alpha})_{ii} = \sum_{j=1}^N (K^S_{\alpha})_{ij}$.\vspace{2pt}
	\item Form the diagonal matrices $\hat D_{ii} = 2\epsilon q_i^{d\beta}$ and $P = \hat D^{1/2} D_{\alpha}^{1/2}$.\vspace{2pt}
	\item Form the symmetric matrix $\hat {\bf L} = P^{-1} K^S_{\alpha} P^{-1} - \hat D^{-1}$. \vspace{2pt}
	\item Find the smallest magnitude eigenvalues $\lambda_j$ and associated eigenvectors $\varphi_j$ of $\hat {\bf L}$.\vspace{2pt}
	\item Normalize the eigenvectors so that $\frac{1}{N}\sum_{l=1}^N \varphi_j(x_l) = 1$.
  \end{enumerate} 

\newpage

\end{document}